\documentclass{article}
 \usepackage[table,xcdraw]{xcolor}
\usepackage[american]{babel}
\usepackage[latin1]{inputenc}
\usepackage{graphicx}
\usepackage{booktabs}
\usepackage{booktabs}
\usepackage{float}
\usepackage{subcaption}
\usepackage{indentfirst}
\usepackage[numbers]{natbib} 
\begin{document}

\title{Ideologically Motivated Biases in a Multiple Issues Opinion Model}


\author{Marcelo V. Maciel \and Andr\'e C. R. Martins\\
	NISC - EACH, Universidade de S\~ao Paulo\\
	Av. Arlindo B\'etio, 1000, S\~ao Paulo, 03828-080, Brazil}


\date{}

\maketitle


\begin{abstract}
	It has been observed people tend to have opinions that are far more internally consistent than it would be reasonable to expect. Here, we study how that observation might emerge from changing how agents trust the opinions of their peers in a model for opinion dynamics with multiple issues. A previous Bayesian inspired opinion model for continuous opinions is extended to include multiple issues. In the original model, agents tended to trust less opinions that were too different from their own. We investigate the properties of the extended model in its natural form. And we also introduce the possibility the trust of the agent might depend not only on the specific issue but on the average opinions over the many issues. By adopting such a ideological point of view, we observe an important decrease in the spread of individual opinions.

\end{abstract}

\section{Introduction}


Here, we will present an opinion dynamics model
\cite{castellanoetal07,galam12a,galametal82,galammoscovici91,sznajd00,deffuantetal00,martins08a}
where the opinions exist over an one-dimensional axis and each agent has a best
estimate on more than one issue over that axis. Describing opinions over a
spatial landscape is an usual way to describe policy alternatives and agents'
preferences. The geometrical properties of the space are usually defined by
mapping from similarity to proximity of the political agents
\cite{downs1957economic, laver2014measuring}. Such a description can capture the
common notion of parties or policies being more ``to the left'' or ``right''. If
they're similar then they're closer \cite{van2005political, miller2015spatial}.
Major opinions, including political ones, tend to be formed not from only one
issue but from how each person feels about a number of them. Locating someone in
a left versus right or liberal versus conservative axis, therefore, requires
inspecting the opinions of that person in not only one but several issues that constitute the agent ideological positioning \cite{benoit2006party}.

For many problems, it makes sense to consider different issues as having components in
more than one single dimension \cite{vicenteetal08b}. However, it is often the case that describing the problem as
one-dimensional can be justified. We can certainly see this as a
first approximation along the most relevant dimension. In that case, the one dimensional case is
just a projection along a direction where variation seems especially important of higher-dimensional problems. From an application point of view, it is usual to find discussions to be simplified over only one main disagreement.  

One traditional way to model this type of scenario is to use continuous opinions
over a fixed interval, as it is done in the Bounded Confidence (BC) models
\cite{deffuantetal00,hegselmannkrause02}. While discrete models
\cite{galametal82,galammoscovici91,sznajd00} can be very useful at describing
choices, they are not easiest way to represent strength of opinion unless a
continuous variable is associated to the choice \cite{martins08a}. Discrete
models also tend to lack a scale where we can compare opinions and decide which
one is more conservative or more liberal.

On the other hand, continuous models are not particularly well suited for
problems involving discrete decisions. As we will not deal with those kinds of
problems here, they are a natural choice. Indeed, continuous opinions models
have been proposed for several different problems on how opinions spread on a
society \cite{deffuantetal02a,weisbuchetal05}, from questions about the spread
of extremism
\cite{amblarddeffuant04,gargiulomazzoni08a,franksetal08a,alizadeh14a,Albi2016,Mai2017}
to other issues such as how different networks
\cite{Kurmyshev2011,Acemoglu2011,Das2014,Hu2017} or the uncertainty of each
agent \cite{deffuant06} might change how agents influence each other.

Here, we will use a continuous opinion model created by Bayesian-like reasoning
\cite{martins08c}, inspired by the Continuous Opinions and Discrete Actions
(CODA) model \cite{martins08a,martins12b}. That model was shown previously
\cite{martins08c} to provide the same qualitative results as BC models. While a
little less simple, the Bayesian basis makes for a clearer interpretation of the
meaning of the variables. That makes extending the model and interpreting new
results simple. That approach is also consistent with a bounded rationality
variant interpretation of the spatial model of political decision making
\cite{humphreys2010spatial,ostrom1998behavioral}.

Variations of how each agent estimate how trustworthy other agents are will also
be introduced. Here, we use a function of trust $p^*$ that plays a role similar
to the threshold value in BC models. While $p^*$ is not a simple discontinuous
cut-off, it is a function of the distance between the opinions of the agent and
the neighbor. If only one issue is debated, that distance is uniquely defined.
However, if we have multiple issues represented on the same one-dimensional
line, it is not clear which distance we should use. That happens because we have
the opinion of agent $i$ on a specific issue $o_i$ and we can compare it to
$o_j$ to estimate how much $i$ trusts $j$. However, as there are several issues
we could also use the mean opinion of $i$ over all its issues and the same goes
for $j$. Therefore, we will study two additional cases: in the first
alternative, we will change \(p^*\) to \(p^{**}\), determined by the distance
between the neighbor and the agent average opinions. In the second alternative,
we will use a \(p^{***}\) calculated from the opinion of the neighbor and the
average opinion of the agent. The idea here is to make the behavior of our
agents closer to what experiments show about human reasoning. We have observed
that our reasoning about political problems can be better described as
ideologically motivated \cite{jostetal03a,taberlodge06a,Claassen2015a}. Indeed,
our opinions tend to come in blocks even when the issues are logically
independent \cite{jervis76a}. Our reasoning abilities seem to exist more to
defend our main point of views \cite{mercier11a,merciersperber11a} and our
cultural identity \cite{kahanetal11} than to find the best answer. In that
context, evaluating others by how they differ from us as a whole, instead of in
each issue, is a model variation worth exploring.
 
\section{The Model}

Here, the population will consist of  \(N\) agents fully connected (an agent $i$ can interact with any other agent $j$). Each agent $i$ will have an opinion $0\leq o_{ik} \leq 1$, where $k=1, \ldots, n$ is a specific issue. We assume each agent opinion about issue $k$ can represented as a value $o_{ik}$ at the range of possible values for $o$s. Agents also have an uncertainty $\sigma$ associated to their average estimate $o$. The uncertainty $\sigma$ could be different for each agent and also updated during the interactions \cite{martins08c}. For the sake of simplicity, however, we will assume the uncertainty $\sigma_i =\sigma$ is identical for (almost) all agents and it does not change. The set of opinions for each agent on all possible issues will define its ideological profile
\(I_i
=
(
(o_{i 1}, \sigma),
\ldots,
(o_{i n}, \sigma)
)
\)
, where \(n\) is the number of issues, \(o\) is
the opinion about the issue and \(\sigma\) is the global  uncertainty~\cite{martins12b}. The arithmetic mean  $x_i$  of agent $i$ opinions in each issue will be called here the ideal point for each agent, that is, it defines the agent ideological position at the
dimension of interest \cite{armstrong2014analyzing}. Obviously
\(
x_i
=
\frac{1}{n}
\sum_{k=1}^{n}
o_{ik}
\).

In order to have agents with initial ideal points well distributed over the
possible range, the initial valued for each \(o_{ik}\) was randomly drawn
using a Beta \(Be(\alpha, \beta)\) with random parameters $\alpha_i$ and
$\beta_i$. Those parameters were drawn for each agent $i$ from the ranges \( (
\alpha \in [1.1, 100], \beta \in [1.1, 100] ) \) by dividing those ranges by
\(N\) equally spaced values and then assigning them to each agent after mixing
the values as in a Latin hypercube sampling~\cite{mckay2000comparison}. That is done to
allow agents with a diverse value of initial ideal points and, at the same time,
to keep the initial \(o_{ik}\)s of each agent $i$ correlated.

While we could have \(\sigma\) as a measurement of each agent uncertainty and
have it evolve with time, here we will keep it as a fixed parameter of the
model. However, a proportion \textit{p$\_$intran} of agents will be stubborn
about one single issue $k$, so that \(\sigma_{ik} = 1e-20\). That behavior is
kept constant as the simulation unfolds, that is $\sigma_{ik}$ is also not
updated by the model. A proportion of agents who are stubborn is introduced here
so we can check if inflexible \cite{galam05b,deffuant2002can,martinsgalam13a}
have a significant impact on the outcomes of the model.



In each iteration of the simulation, two procedures are applied: the
opinion update through social influence, and a random opinion update (noise). In
the social influence procedure we draw two agents, \(i\) and \(j\), from the
population, where \(i\) will observe \(j\) opinion about a randomly drawn issue \(k
\in (1 , \ldots, n)\). 
Agent \(i\)
updates its opinion (\(o_{ik}\)) following an approximate Bayesian rule. That rule is obtained by assuming each agent has a
Normal prior \(f_i(\theta) = \frac{1}{\sqrt{2 \pi} \sigma_i} e^{- \frac{(\theta
- o_i )^2}{2 \sigma_i}} \) compatible with its parameters $o_{i}$ and $\sigma_{i}$, where the index related to the issue $k$ was omitted for the sake of simplifying the equation.  The agents also assume a mixture likelihood where there is an initial chance
\(p\), updated to \(p^*\) that the agent \(j\) has information and a chance
\(1-p\) it knows nothing and its opinion is just a random non-informative draw. That is, the likelhood is given by
 \( f(o_j|\theta) = p N(\theta,\sigma_j^2) + (1-p)U(0,1) \). While a full Bayesian treatment would produce a posterior distribution that is a mixture of two normals, here we will assume each agent only updates its expected value and it does not carry the full posterior information to the next iteration. That leads to the following update rule for the expected value $o_{i,k}$ \cite{martins12b}:

  \begin{equation}\label{eq:oupdate}
    o_{ik}(t+1) =
    p^{*}
    \frac{o_{ik}(t) + o_{jk}(t) }{2}
    +
    (1 - p^{*})
    o_{ik}(t),
  \end{equation}
where  $p^{*}$ is given by
  \begin{equation}\label{eq:pstar}
   p^{*}
    =
  \frac{
      p \frac{1}{\sqrt{2 \pi} \sigma_i}
      e^{- \frac{ (\Delta_{ij})^2}{2 \sigma_i^2}}
    }{
      p
      \frac{1}{\sqrt{2 \pi} \sigma_i}
    e^{- \frac{ ( \Delta_{ij})^2}{2 \sigma_i^2}}
    +
    (1 - p)
  }.
  \end{equation}
Here \(\Delta_{ij} = o_{ik} (t) - o_{jk} (t)\) is the distance between the opinions on the $k$ issue. $\Delta_{ij}$ plays a
similar role to the threshold parameter in the Bounded Confidence models by making distant opinions less influential. As $\Delta_{ij}$ increases, it is easy to see that Equation\ref{eq:pstar} causes  $p^{*}$ to tend to zero. And, as that happens, the weights in average update Equation~\ref{eq:oupdate} change so that the previous value $ o_{i,k}(t)$ remains almost unchanged.

The threshold role of $\Delta_{ij}$ suggests we might change its definition to check how it might better reflect the actual behavior of humans. In particular, people tend to trust better those who have a similar ideology. That means trust might not depend on the specific issue alone, but on the average over all issues. While that is not the model we obtain from applying Bayes rule, such a change makes sense
as an attempt to have a better model. In order to check that possibility we will also implement two other cases by changing the way $\Delta_{ij}$ is calculated. In the first variation, $p^*$ will be substituted by \(p^{**}\), where \(p^{**}\) means 
 \(\Delta_{ij} = x_i(t) - x_j(t) \), that is, agent $i$ will observe the average
 ideological position of $j$ in order to estimate its trust. That assumed $i$
 has more information than only the value $o_{jk}(t)$. To check what happens
 when only $o_{jk}(t)$ can be observed by agent $i$, we will introduce a second variant case where $i$ compares $o_{jk}(t)$ to its own average. That is,  \(\Delta_{ij} = x_{i}(t) - o_{jk}(t)\) and we represent that case by \(p^{***}\).

In order to better understand the model dynamics we also introduce noise as the second procedure. At each iteration, 
another agent \(i\) is randomly chosen and its opinion changed due to random noise. That is, $i$ opinion becomes \(
o_{ik}(t+1) = o_{ik}(t) + r \) where \(r\) is drawn from a Normal
distribution with mean 0 and standard deviation \(\rho\). If agent \(i\) is intransigent in
issue \(k\) it won't its \(o_{ik}\) opinion when chosen by
the noise algorithm. Moreover, if \(o_{ik}(t) + r > 1\) then \( o_{ik}(t+1) =
1\). Likewise, if \(o_{ik}(t) + r < 0 \) then \( o_{ik}(t+1) = 0\). Noise
is introduced here as a way of accounting for the effect of factors not related to the modeled
social influence \cite{flache2017} and it is interesting to verify if small noises can lead to drastic changes in
systemic properties \cite{macy2015signal}.

  \section{Model Results}

  To better understand the model behavior, we ran simulations using as range for the parameters the values:  

  \begin{table}[H]
    \centering
\begin{tabular}{@{}|l|l|l|l|l|l|@{}}
\toprule
\rowcolor[HTML]{EFEFEF} 
$\sigma$ & $n$ & $p$ & $p\_intran$ & $N$ & $\rho$ \\ \midrule
$[0.01, 0.5]$ & $[1, 10]$  & $[0.1, 0.99]$ & $[0.0, 0.1]$ & $[500, 5000]$ & $(0.0, 0.1]$ \\ \bottomrule
\end{tabular}
\caption{Parameters' Bounds}
\end{table}

\begin{figure}[H]
  \centering \captionsetup{justification=centering,margin=2cm}
    
  \begin{subfigure}[h]{0.85\textwidth}
    \includegraphics[width=\textwidth]{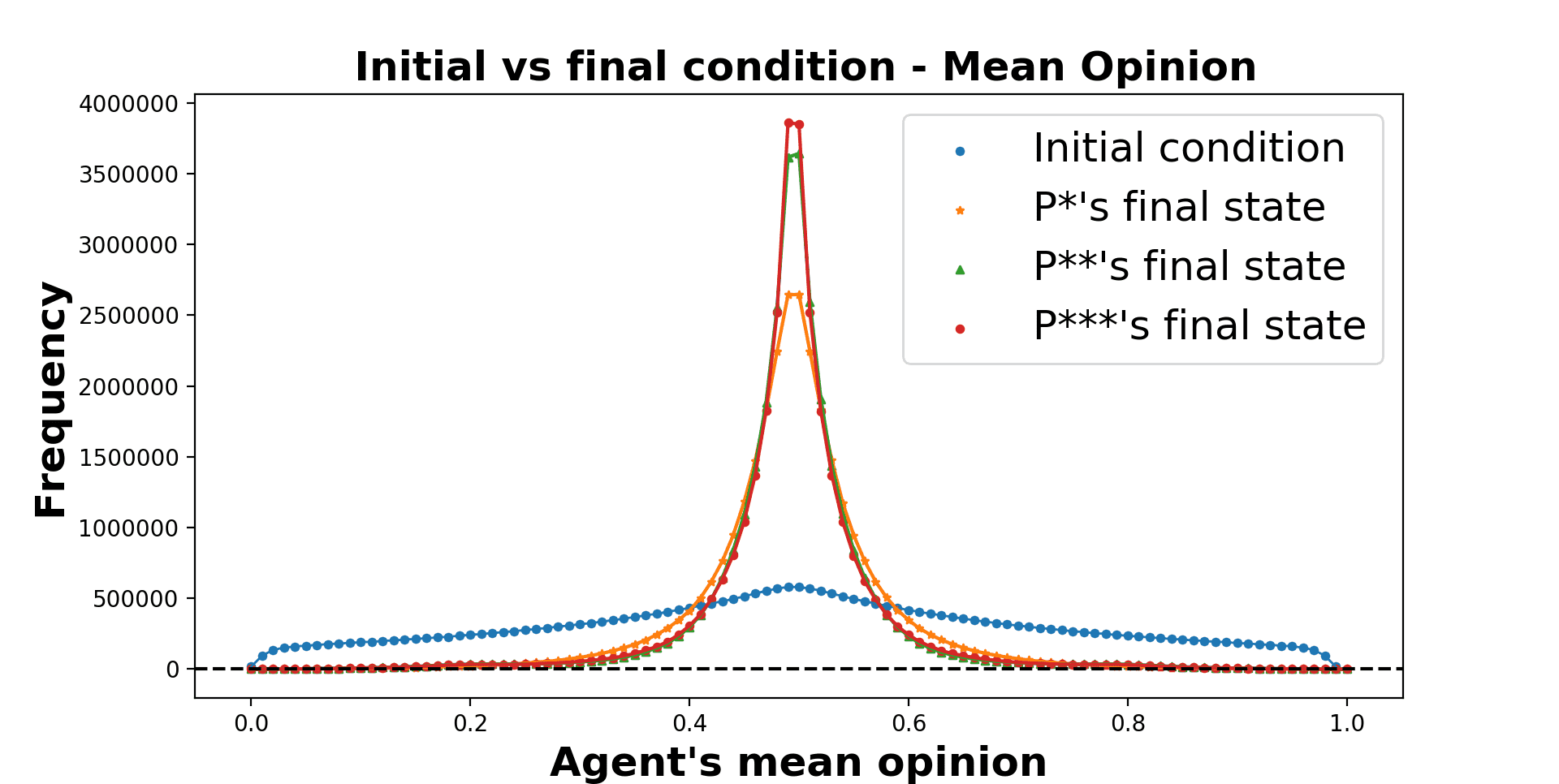}
    \caption{Initial x Final state: \\ Distribution of mean opinions}
  \end{subfigure}
  \begin{subfigure}[b]{0.85\textwidth}
    \includegraphics[width=\textwidth]{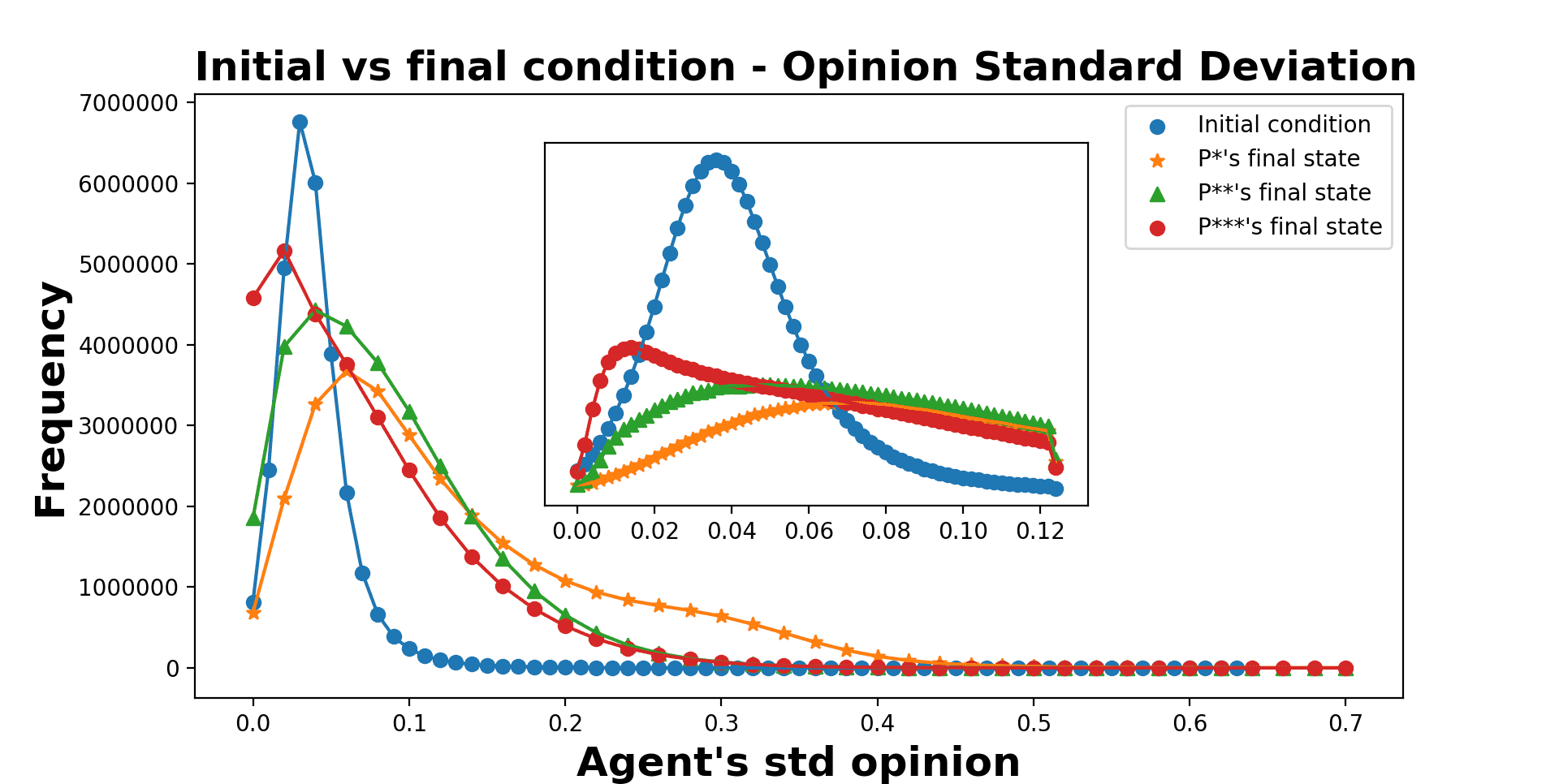}
    \caption{Initial x Final state: \\ Distribution of the standard deviations of opinions}
  \end{subfigure}
  \caption{Difference between initial and final opinion distributions for 60.000
    parameterizations. In all cases, \(N = 500\).}
  \label{fig:std}
\end{figure}

The parameter space was explored by two sweeps of its parameters: one sampling
of 70,000 times using quasi-random low-discrepancy sequences on all parameters
\cite{saltelli2008global}, that generate evenly spaced points, and another of
60,000 times keeping \(N=500\) so that we can compare different runs. An
interesting question in any opinion dynamics model is if agents can reach
consensus, if they diverge, or something between those two states. That can be observed by comparing the initial mean and standard
deviation of each agent's opinions. Figure \ref{fig:std} shows histograms for
those variables at the initial condition and also for the final opinions corresponding to the (\(p^{*}, p^{**}, p^{***}\)) cases. The upper graphic shows the distribution of the mean opinions $x_i$, and the lower one the distribution of the standard deviation of the opinions of each agent $s_i^2==
\frac{1}{n}
\sum_{k=1}^{n}
(o_{ik}-x_i)^2$.

The histograms show that, in general, opinions have a tendency to move towards
the middle value. This seems to suggest most parameter values lead to consensus
or near it. However, consensus is not always achieved, as shown by the tails of
the mean distribution at Figure \ref{fig:std}, even if the final distribution
shows that extreme opinions become less common. And we can also observe in the
distribution of standard deviations $s_i$ that the opinions of each agent on
each issue tend to become more diverse, as, for most cases, $s_i$ seems to
increase from its initial values. The only exception is the $p^{***}$ scenario.
There, we still see an increase in the frequency of large values of $s_i$. But
there are also many cases where there is a strong tendency for $s_i$ to become
smaller. Or, in other words, in several scenarios, the agents opinions on the
issues tended to gather much closer to their own mean opinions $x_i$ as a
consequence of the dynamics.

The cases where the average opinions remain a little spread, while rare,
correspond to scenarios where bi-partisanship survives, with agents opinions
surviving at both extreme positions. This suggests that, in general, the model
can be described as one with similarity biased influence \cite{flache2017}. This
tendency to consensus seems to be a little weaker at the \(p^*\) case when
compared to the two other cases, $p^{**}$ and $p^{***}$.

Figure \ref{fig:std} (b) also shows that in all cases (\(p^{*}, p^{**},
p^{***}\)) the standard deviation of the opinions distribution becomes more
spread. While there are situations where $s_i$ tends to become smaller,
signaling the agent opinions become closer to its mean, the reverse also
happens. The tails for larger values of $s_i$ show that the intercation with
other agents quite often led to a more diverse set of opinions over the
different issues. As a matter of fact, except for the $p^{***}$ scenarios, a
larger internal spread than in the initial conditions seem to be the usual
result. For $p^{***}$, while we also observe that strenghtening of the spread
for many parameter values, we also see that an increase in cases with small
$s_i$. That is, the dynamics can lead to stronger internal consistency.

The distributions for the mean and standard deviation might look contradictory
at a first glance. But they are information about different quantities. A
tendency of the $x_i$ towards central value while $s_i$ increases is simple to
understand. It just shows that, while opinions on different issues might be
spreading more, including towards more extreme values, the mean of every issue
opinion of the agents show a clear central tendency. That the mean would have a
central tendency should not be seen as a surprise. But that does not necessarily
correspond to what happens to the agents  opinions on specific issues $o_{ik}$.

    \begin{figure}[H]
  \centering
    \begin{subfigure}[b]{0.49\textwidth}
      \includegraphics[width=\textwidth]{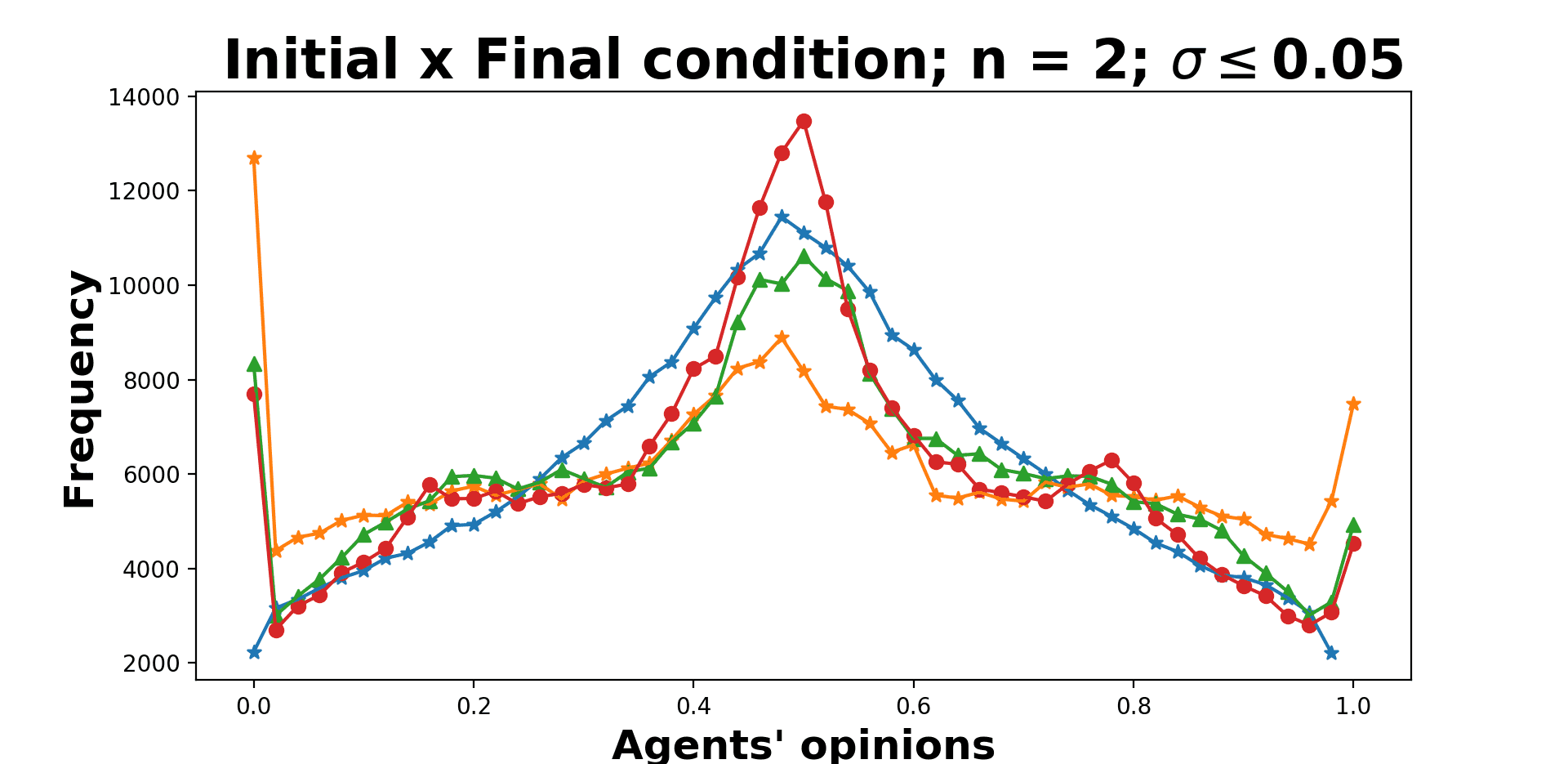}
    \end{subfigure}
         \begin{subfigure}[b]{0.49\textwidth}
      \includegraphics[width=\textwidth]{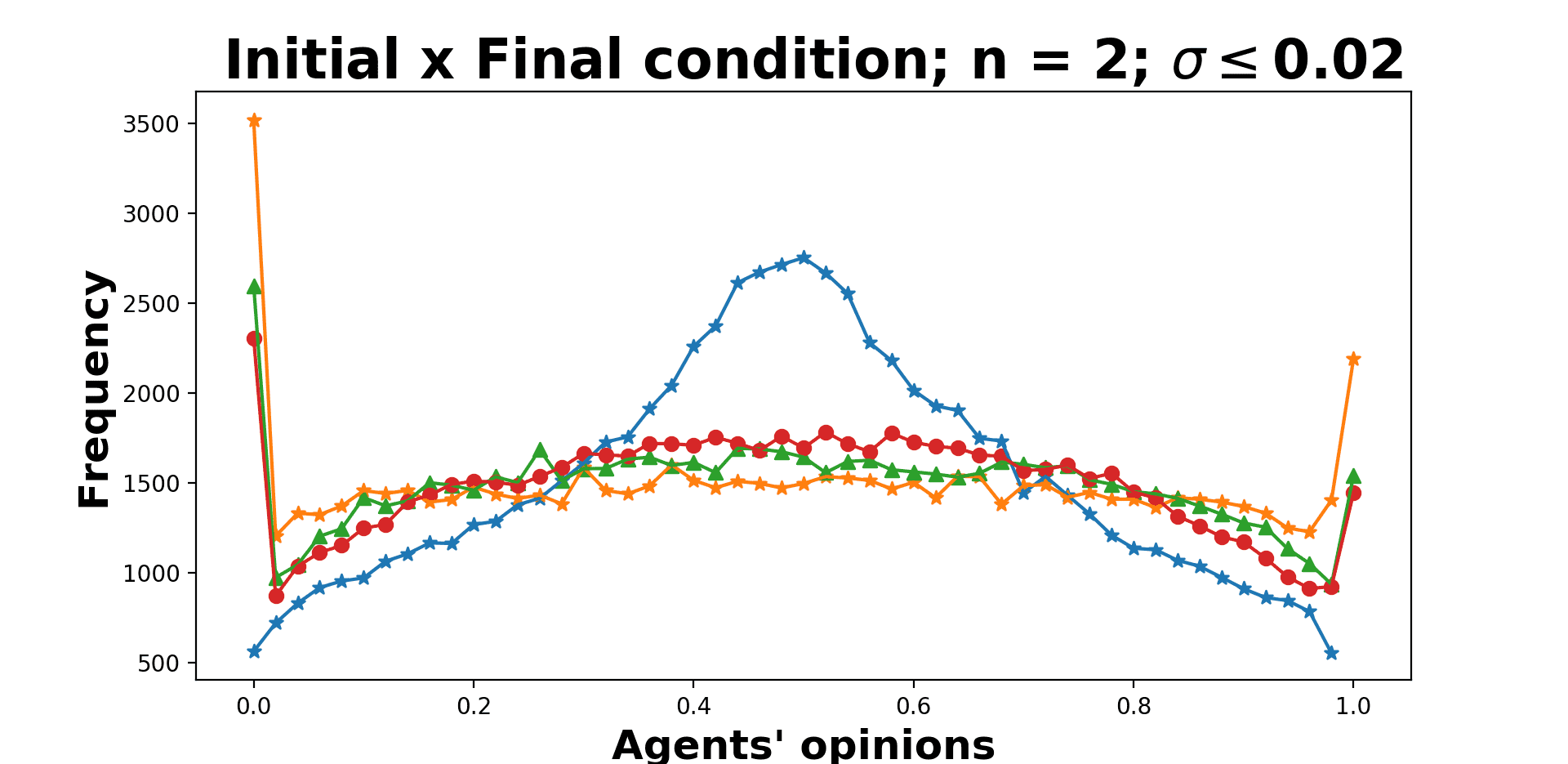}
    \end{subfigure}
     
    \begin{subfigure}[b]{0.49\textwidth}
      \includegraphics[width=\textwidth]{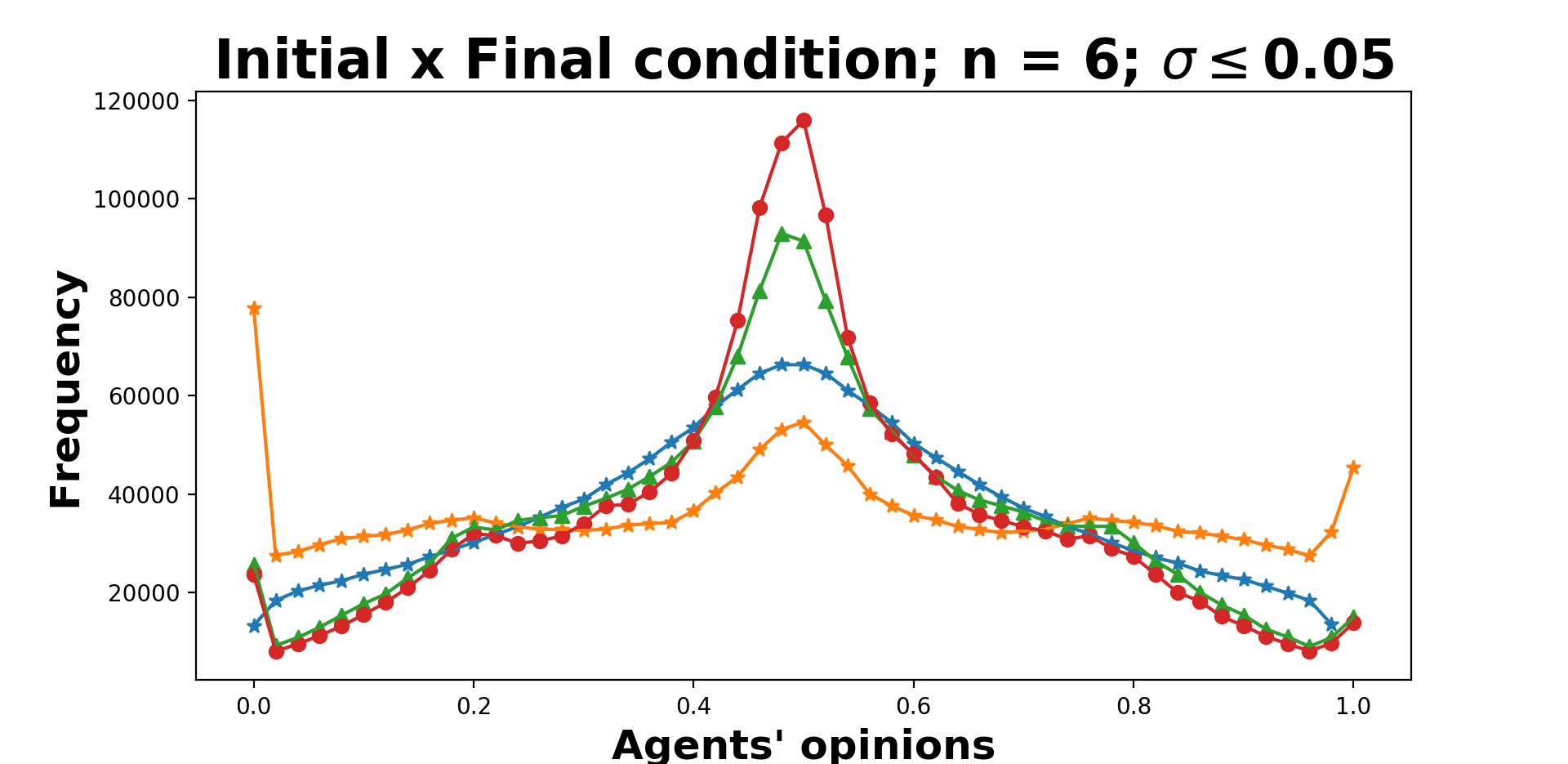}
    \end{subfigure}
    \begin{subfigure}[b]{0.49\textwidth}
      \includegraphics[width=\textwidth]{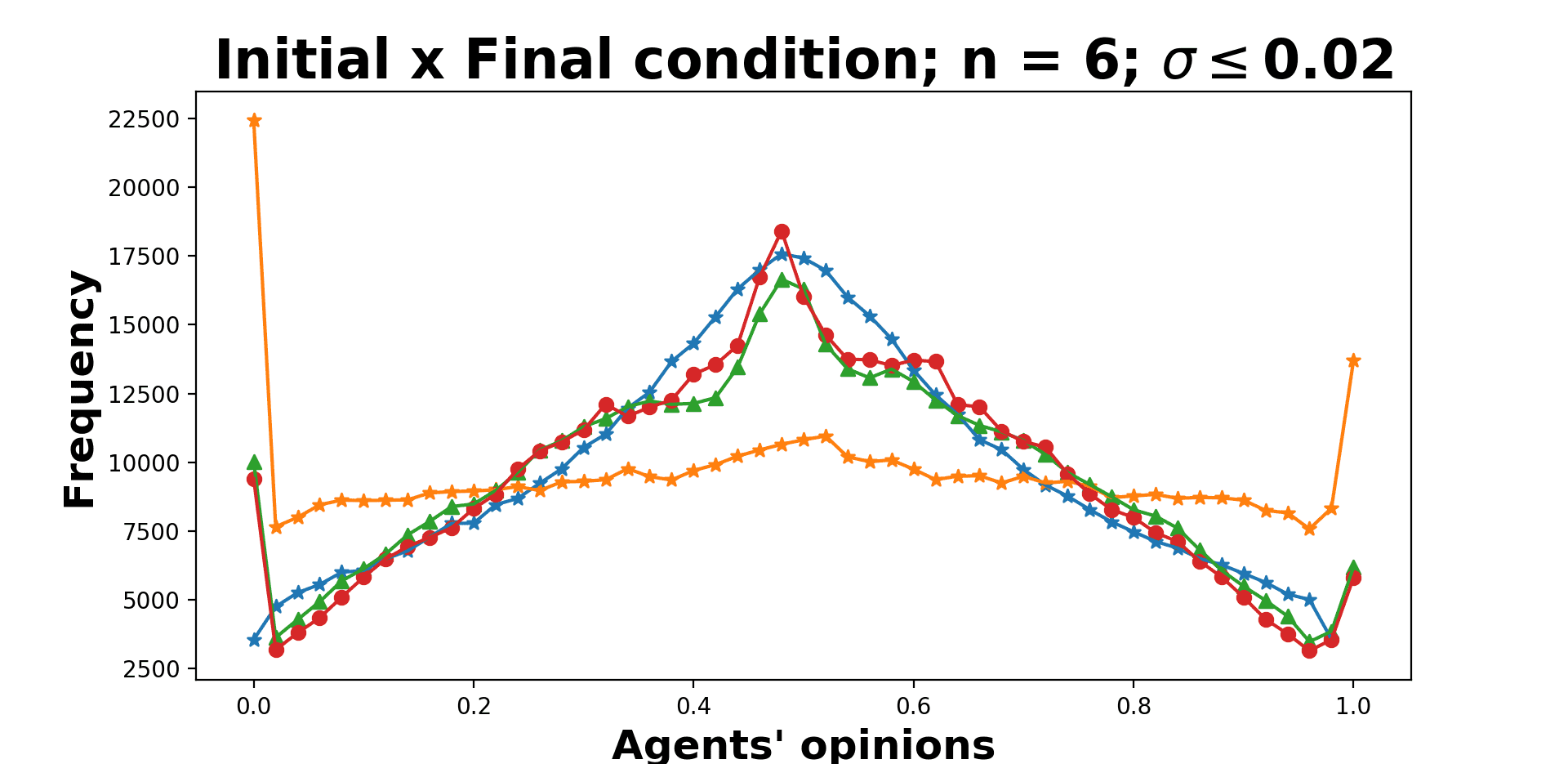}
    \end{subfigure}
     \begin{subfigure}[b]{0.49\textwidth}
       \includegraphics[width=\textwidth]{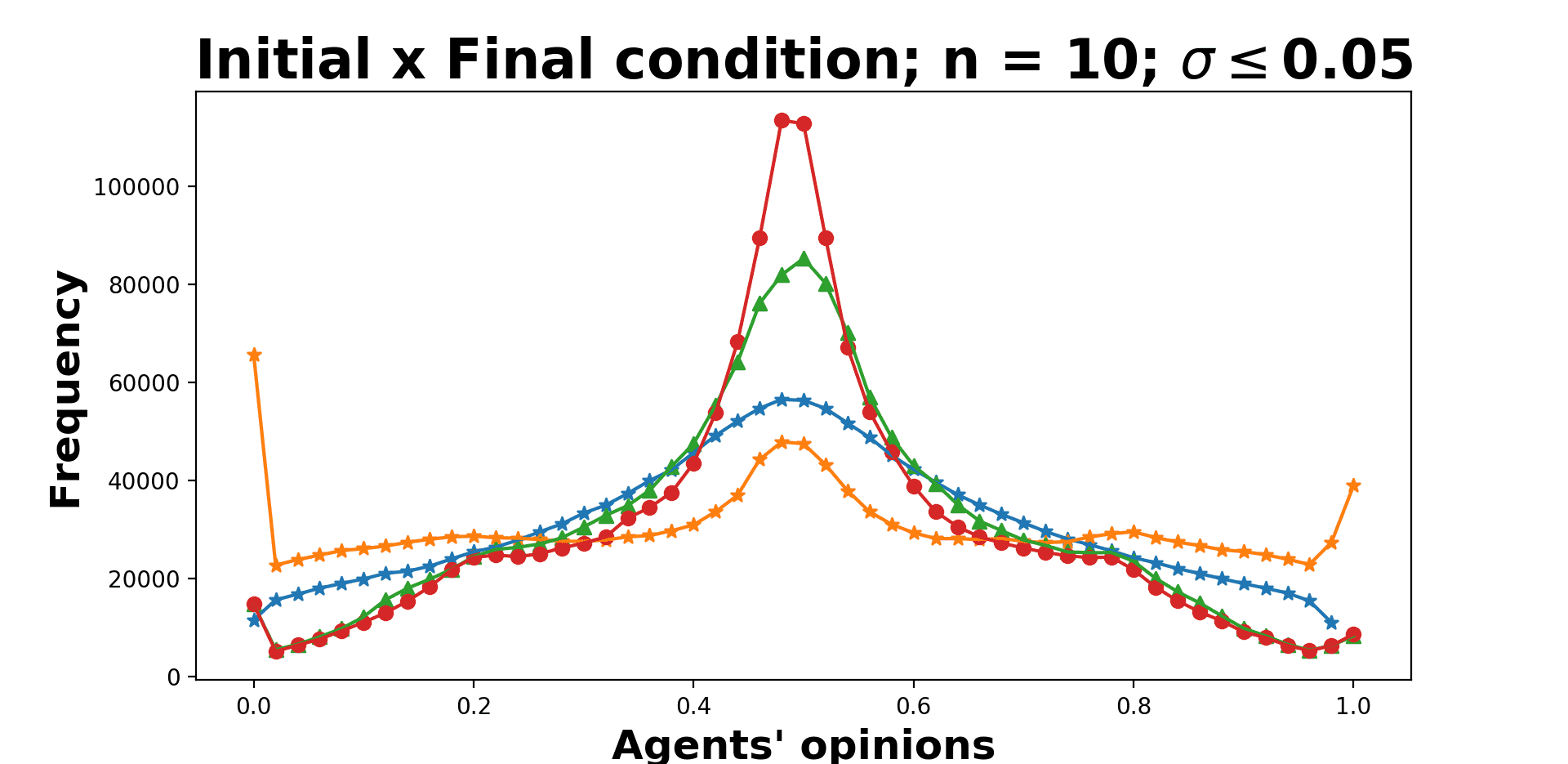}
     \end{subfigure}
     \begin{subfigure}[b]{0.49\textwidth}
       \includegraphics[width=\textwidth]{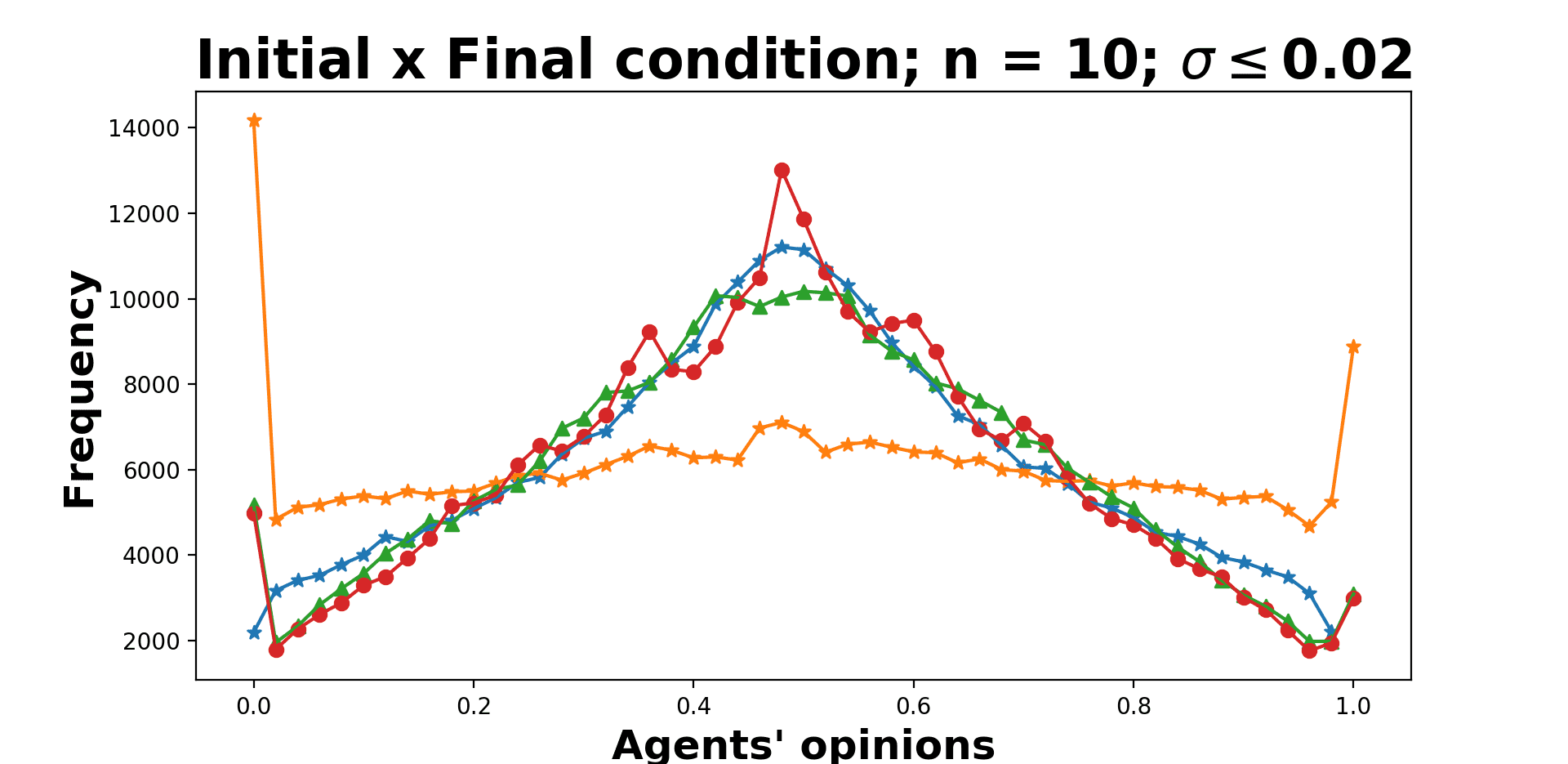}
     \end{subfigure}
     \begin{subfigure}[b]{0.49\textwidth}
       \center
       \includegraphics[scale=0.1]{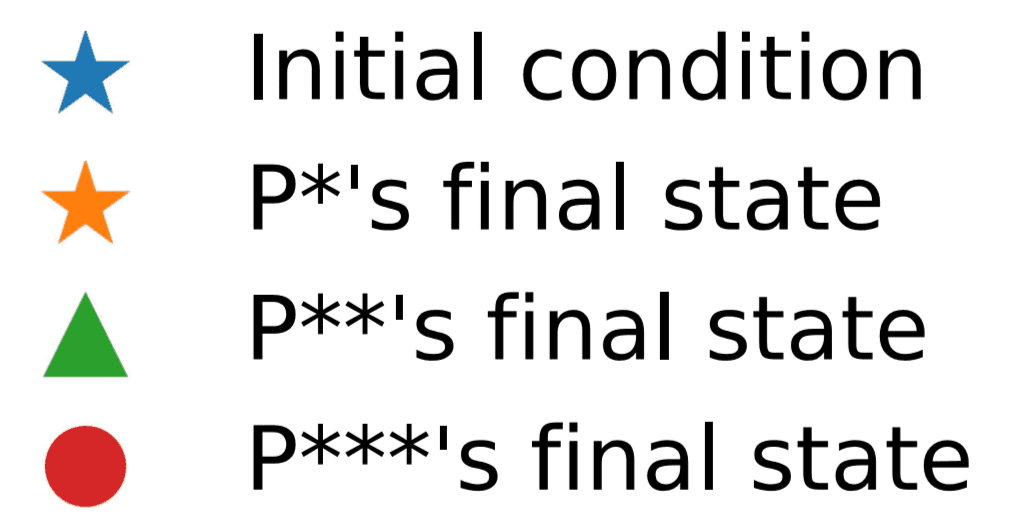}
     \end{subfigure}

	\caption{Histogram for the final distribution of values of $o_{ik}$. Initial distribution is also shown for comparison.}
	\label{fig:oik}
\end{figure}

The behavior of the opinions about specific issues $o_{ik}$ values when we have
small values of $\sigma$ can be observed in the Figure~\ref{fig:oik}. The
graphics show separate cases for different number $n$ of issues and distinct
small values of $\sigma$. The stronger central tendency. While the most
important peak is still around 0.5, the distribution of values is now much more
spread. That reflects the fact that, as $\sigma$ decreases, consensus becomes
much harder and a fraction of the agents find agreement at other values. Indeed,
in every graphic, for the $p^*$ scenario, central opinions become less common
than in initial conditions, while more extreme values become predominant,
showing a clear tendency away from consensus. The same is not true for the more
centralizing versions, $p^{**}$ and $p^{***}$, that seems to show a more varied
behavior. For $\sigma \leq 0.05$, both cases show some tendency to consensus. As
for $\sigma \leq 0.02$, we see very small changes in the initial distribution,
when $n=6$ or $n=10$, and a clear drive to disagreement when $n=2$. In all
scenarios, however, it is clear that the tendency to disagreement is weaker when
update is done following the $p^{**}$ and $p^{***}$ rules than what we observe
for $p^*$.

The histograms, however, don't show the whole story of which parameters influence the system behavior. With that in mind we performed a
Sobol sensitivity analysis \cite{saltelli2000sensitivity} using as outcome both the observed
standard deviation of agent's mean opinions \(Ystd ^2==
\frac{1}{n}
\sum_{k=1}^{n}
(x_{i}-\bar{x_i})^2  \) and the standard deviation $S_s$ of the collection of $N$ internal standard deviations $s_i$. The Sobol indexes decompose the impact of parameters on
the variance of the output. The higher the value of index the bigger the impact
of the parameter on the output. First order Sobol indexes include linear and
non-linear contributions of the parameters, while total Sobol indexes also
include all the interaction effects between parameters
\cite{ten2016sensitivity}. If there are only three parameters, the total effect
\(S_{T1}\) of the first parameter \(X_1\) is given by the equation \(S_{T1} =
S_1 + S_{12} + S_{13} + S_{123}\) where \(S_i = \frac{V[E(Y|X_i)]}{V(Y)}\);
\(S_{12}\) is the impact at the variance of the output \(Y\) of the interaction
of \(X_{1}\) and \(X_{2}\); that is, their combined effect minus their first
order effects: \(S_{ij}\) = \(\frac{V[E(Y|X_i,X_,)] - V[E(Y|X_i)] -
  V[E(Y|X_j)]}{V(Y)} \) \cite{saltelli2008global}. For our simulations, we
obtain the estimates:

    \begin{figure}[H]
  \centering
    \begin{subfigure}[b]{0.48\textwidth}
      \includegraphics[width=\textwidth]{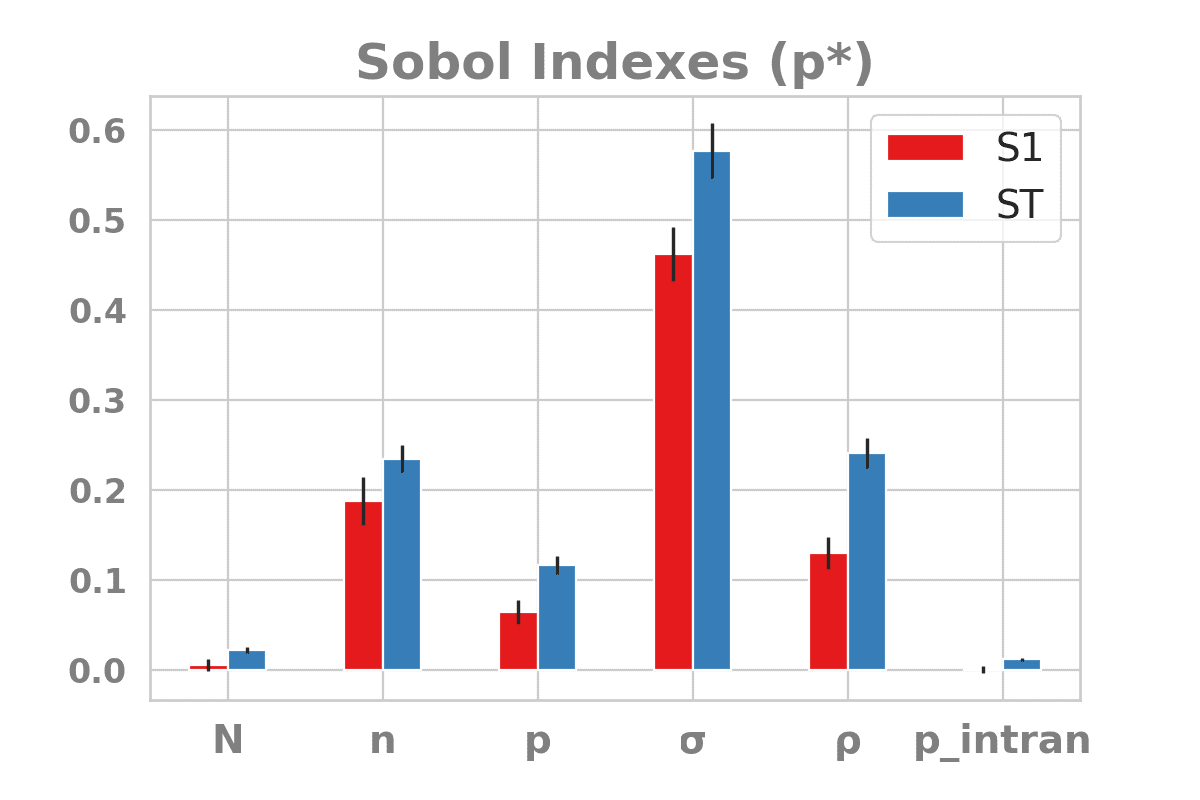}
    \end{subfigure}
         \begin{subfigure}[b]{0.48\textwidth}
      \includegraphics[width=\textwidth]{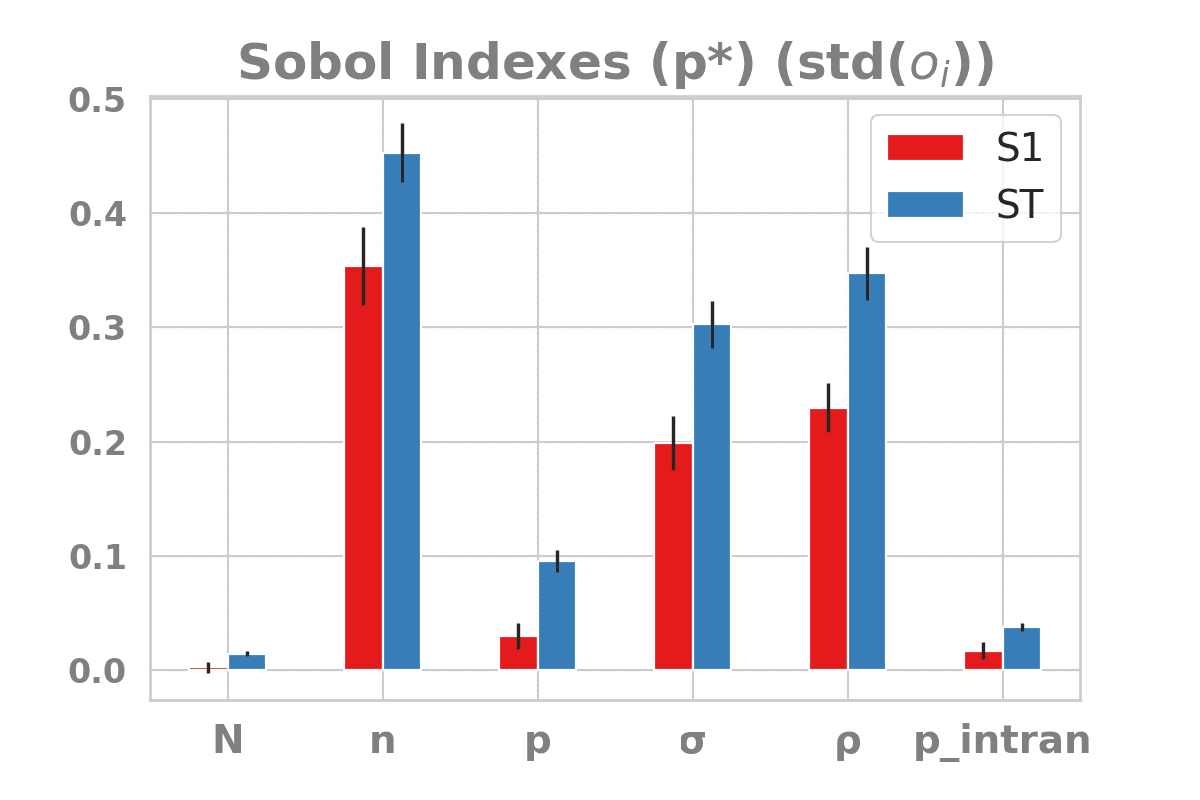}
    \end{subfigure}
    \begin{subfigure}[b]{0.48\textwidth}
      \includegraphics[width=\textwidth]{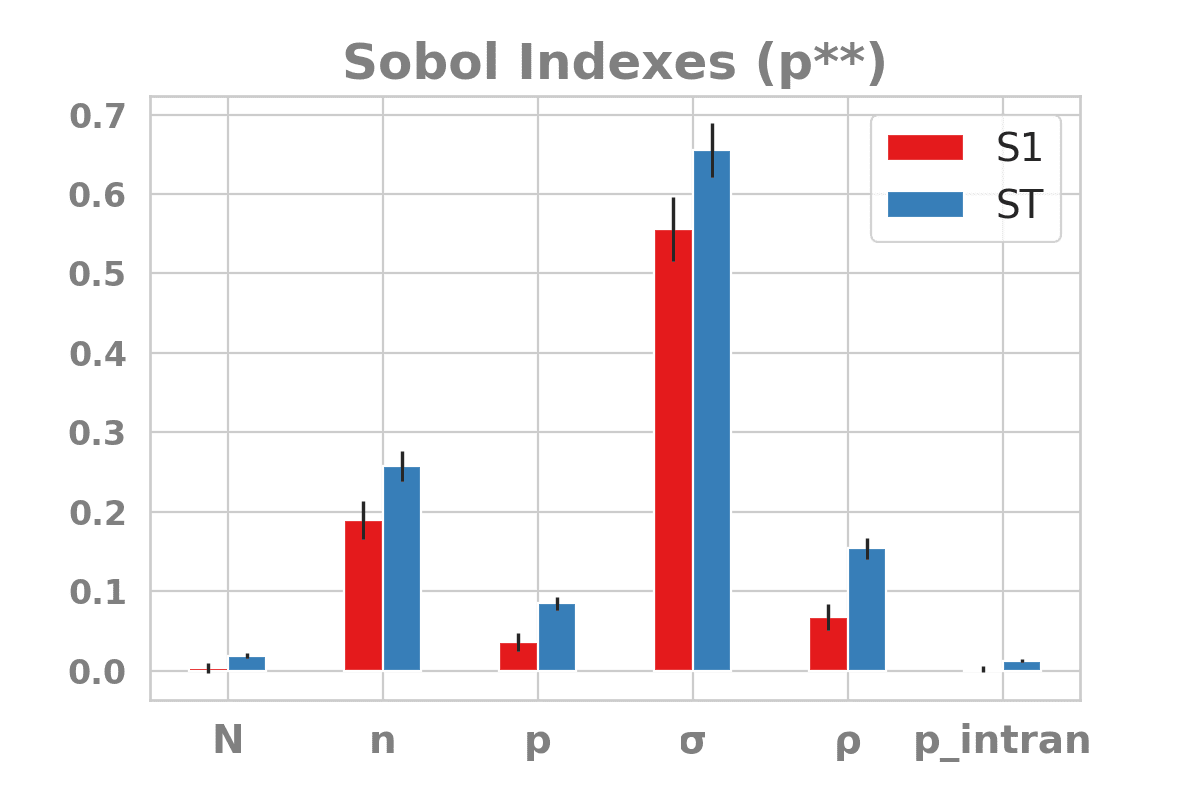}
     \end{subfigure}
    \begin{subfigure}[b]{0.48\textwidth}
      \includegraphics[width=\textwidth]{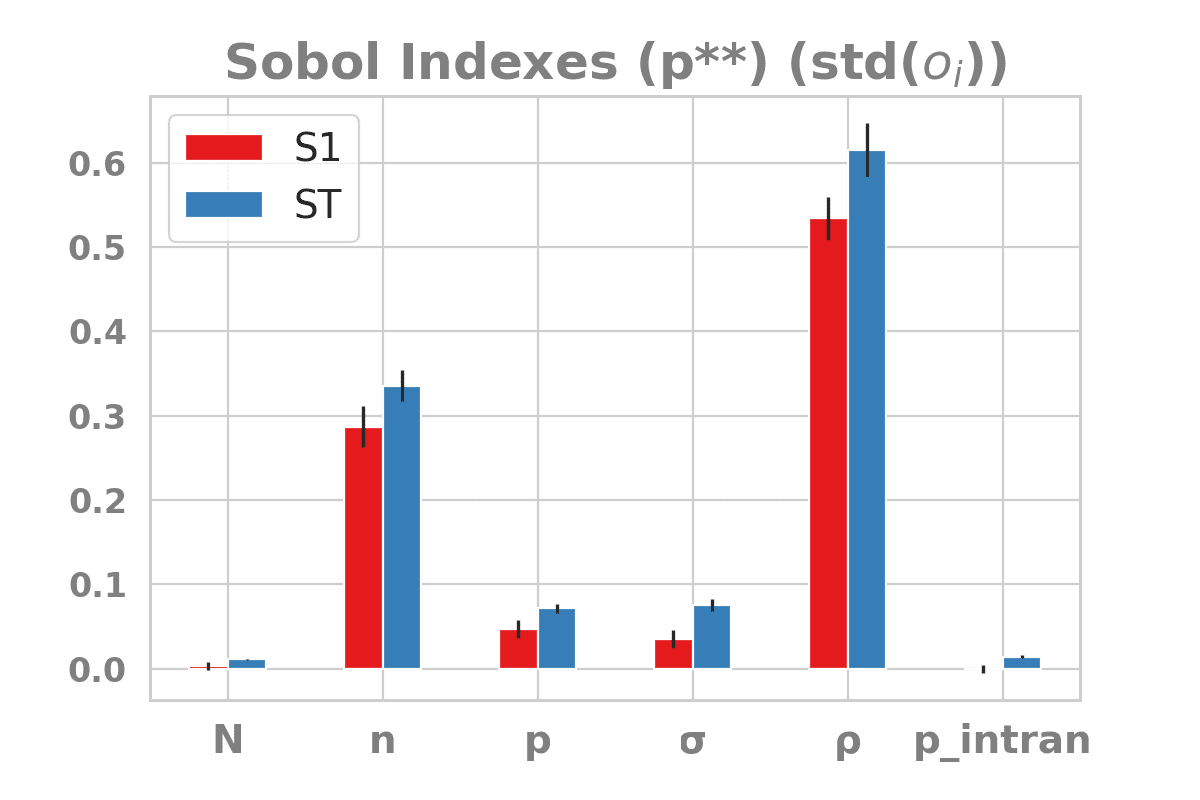}
     \end{subfigure}
     \begin{subfigure}[b]{0.48\textwidth}
       \includegraphics[width=\textwidth]{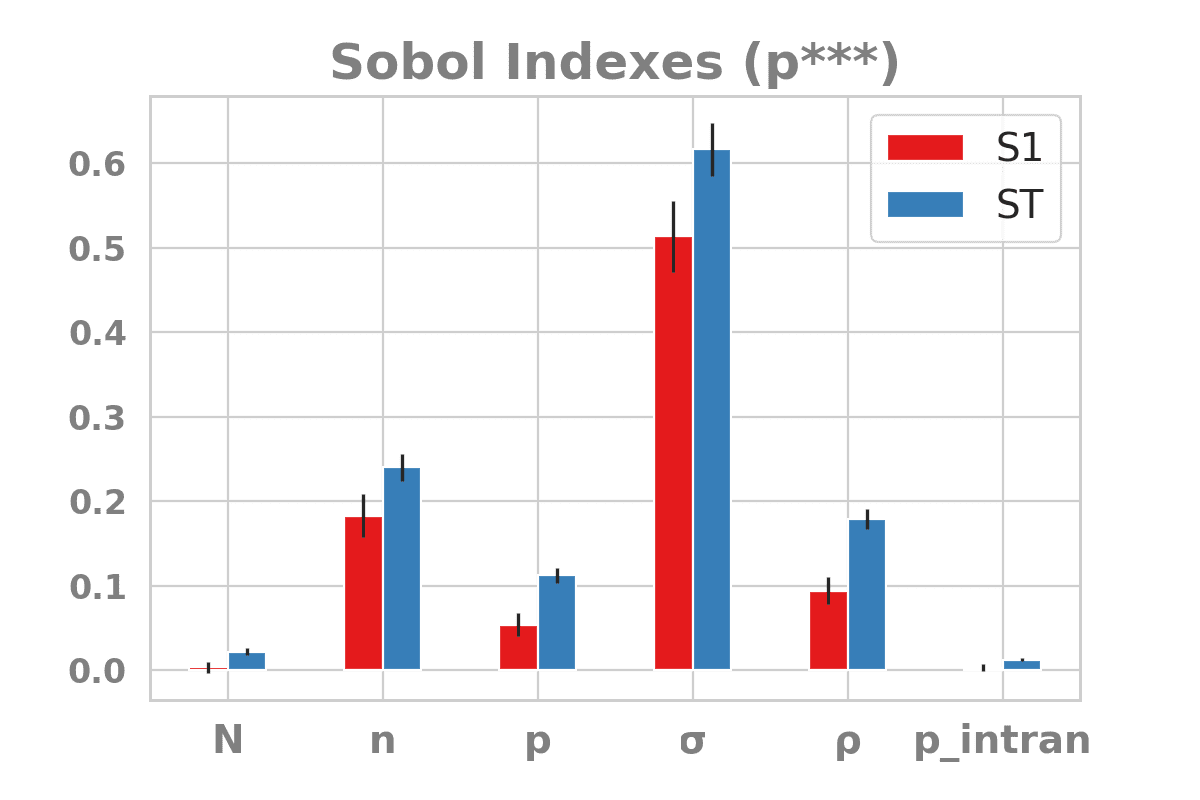}
     \end{subfigure}
     \begin{subfigure}[b]{0.48\textwidth}
       \includegraphics[width=\textwidth]{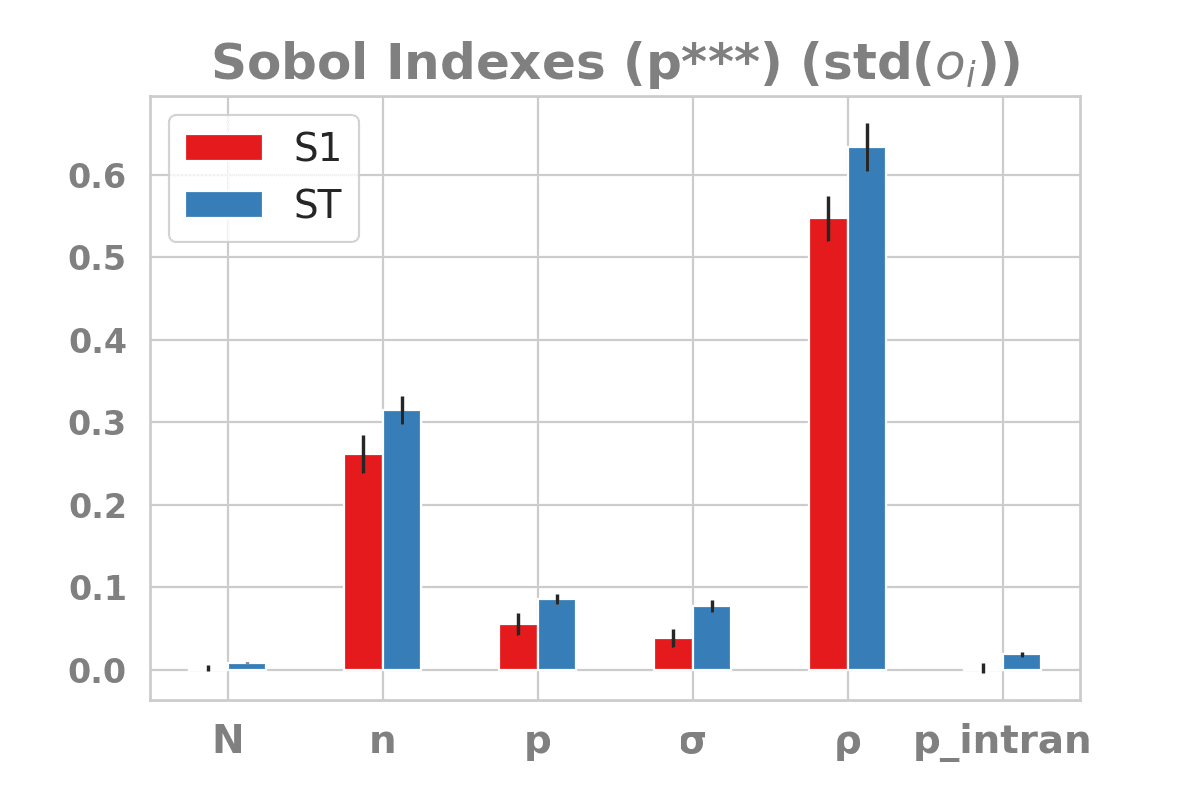}
     \end{subfigure}
     \caption{Sobol Indexes for the three cases (\(p^{*}, p^{**}, p^{***}\)). First column shows the analysis for $Ystd$ and the second column, for $S_s$.}
     \label{fig:sobol} 
    \end{figure}

    The sensitivity analysis in Figure\ref{fig:sobol} shows that \(\sigma, n \),
    and \(\rho\) have the most influence on the final values of $Ystd$ and
    $S_s$. $p$ seems to still have some smaller influence and both $N$ and
    $p_{intran}$ seem to make no difference for the range of values we used. It
    is also interesting to notice that the same general behavior appears in the
    three scenarios for $Ystd$. \(\sigma\) being the parameter that explains
    most of the variance was expected and it is consistent with
    \citep{martins12b}. It is interesting to see that two of the new parameters,
    the number of issues $n$ and the noise \(\rho\), also play an important role
    in explaining the total variance of $Ystd$.
    
    When we look at the standard deviation of the agent's opinions, $S_s$,
    however, we see that \(\sigma\) is the most relevant parameter only for the
    $p^*$ case. For $p^{**}$ and $p^{***}$, however, \(\sigma\) seems to matter
    very little, basically as much as $p$. And the noise $\rho$ seems to have
    most of the impact on $S_s$. This change is not so hard to understand. As we
    are speaking of how much change we observe in the standard deviations of
    each agent opinions $s_i$, there is a fundamental difference between the
    $p^*$ scenario and the $p^{**}$ and $p^{***}$ ones. On the first one, the
    opinion of $i$ in each issue does not depend on its own opinion on other
    issues. However, as in both $p^{**}$ and $p^{***}$ agent $i$ uses its own
    mean to decide own much to trust other agents, it is natural $i$ opinions
    will tend to a mean value. So, while for $p^*$, $s_i$ is driven by how much
    the opinions get spread in general and, therefore, by \(\sigma\) , $p^{**}$
    and $p^{***}$ scenarios suggest that the tendency to keep consistent
    opinions predominates and the variation in the final results is driven
    mostly by the noise. This is compatible with what we observed Figure
    \ref{fig:std} for the distribution of $s_i$. There, we had that the $p^*$
    showed it was common to observe agents with very little internal
    consistency, equivalent to large amounts of $s_i$. The tail for high $s_i$,
    however, was significantly smaller for $p^{**}$ and $p^{***}$, with
    $p^{***}$ even showing cases where very small $s_i$ became more frequent
    than in the initial distribution.

\begin{figure}[H]
  \centering
  
    \begin{subfigure}[b]{0.48\textwidth}
      \includegraphics[width=\textwidth]{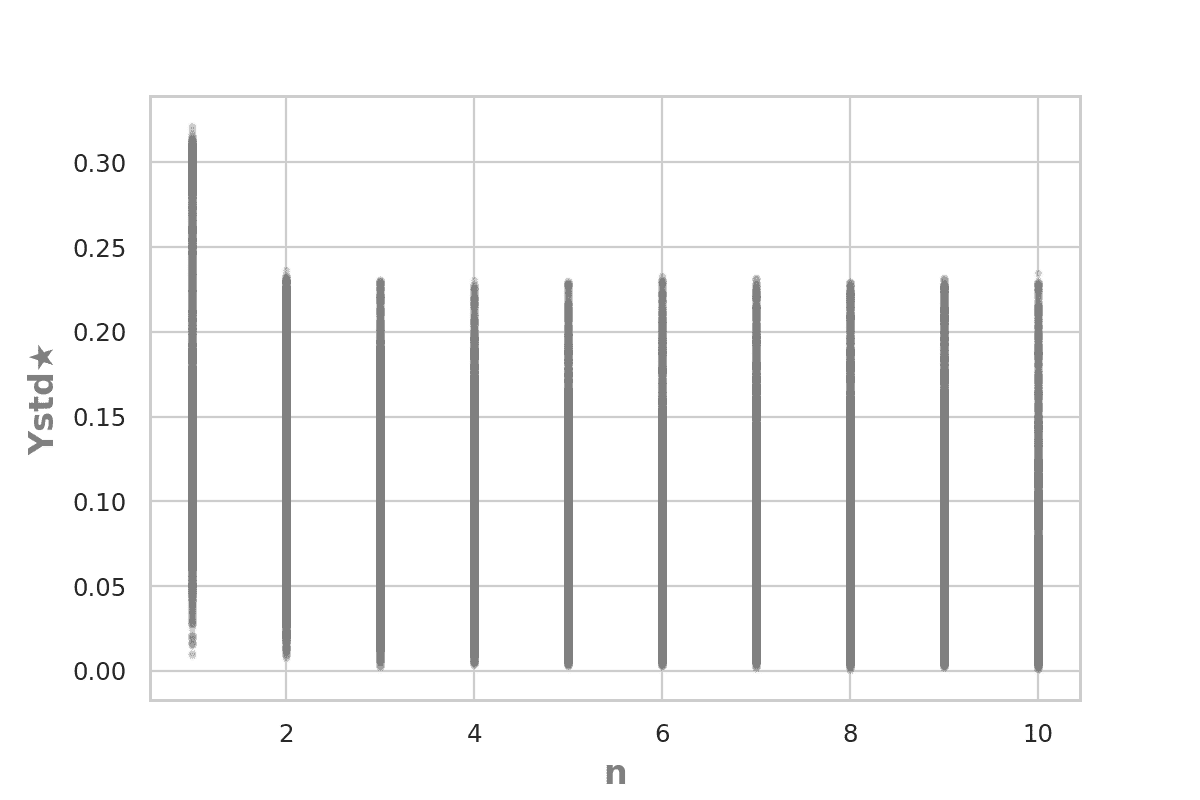}
    \end{subfigure}
    \begin{subfigure}[b]{0.48\textwidth}
      \includegraphics[width=\textwidth]{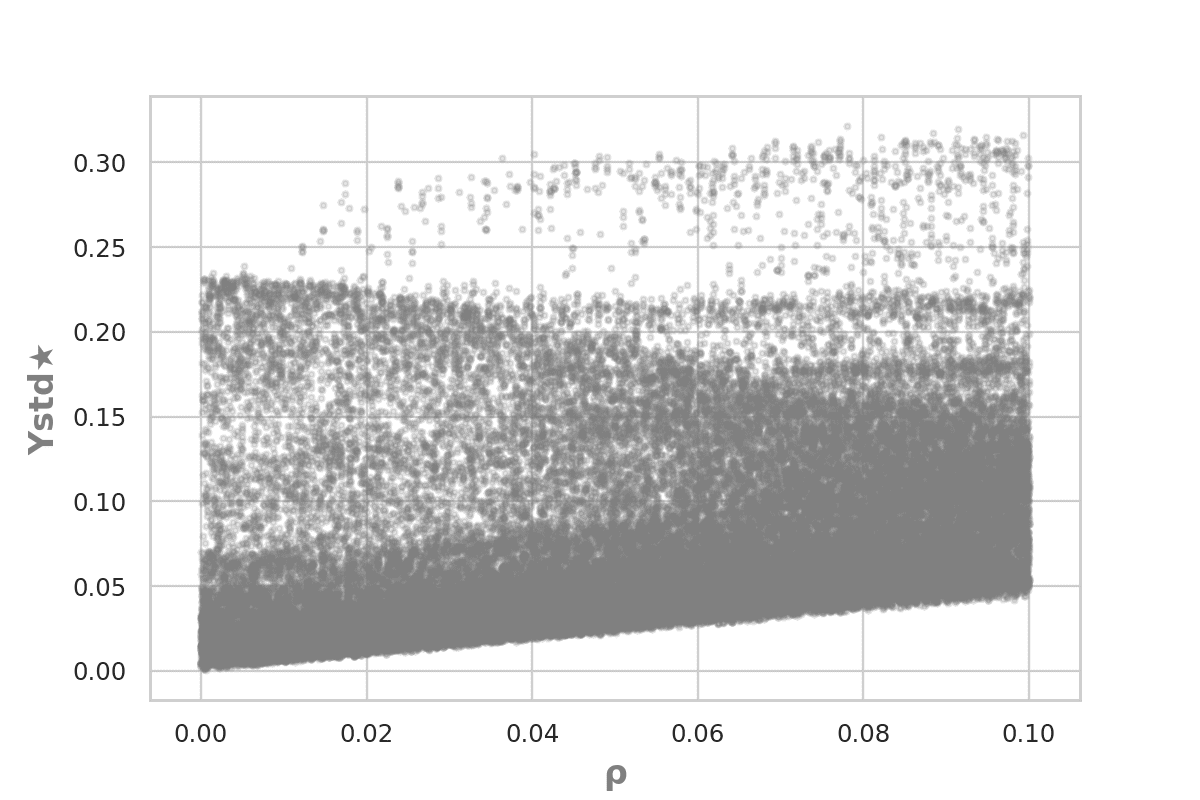}
     \end{subfigure}
     \begin{subfigure}[b]{0.5\textwidth}
       \includegraphics[width=\textwidth]{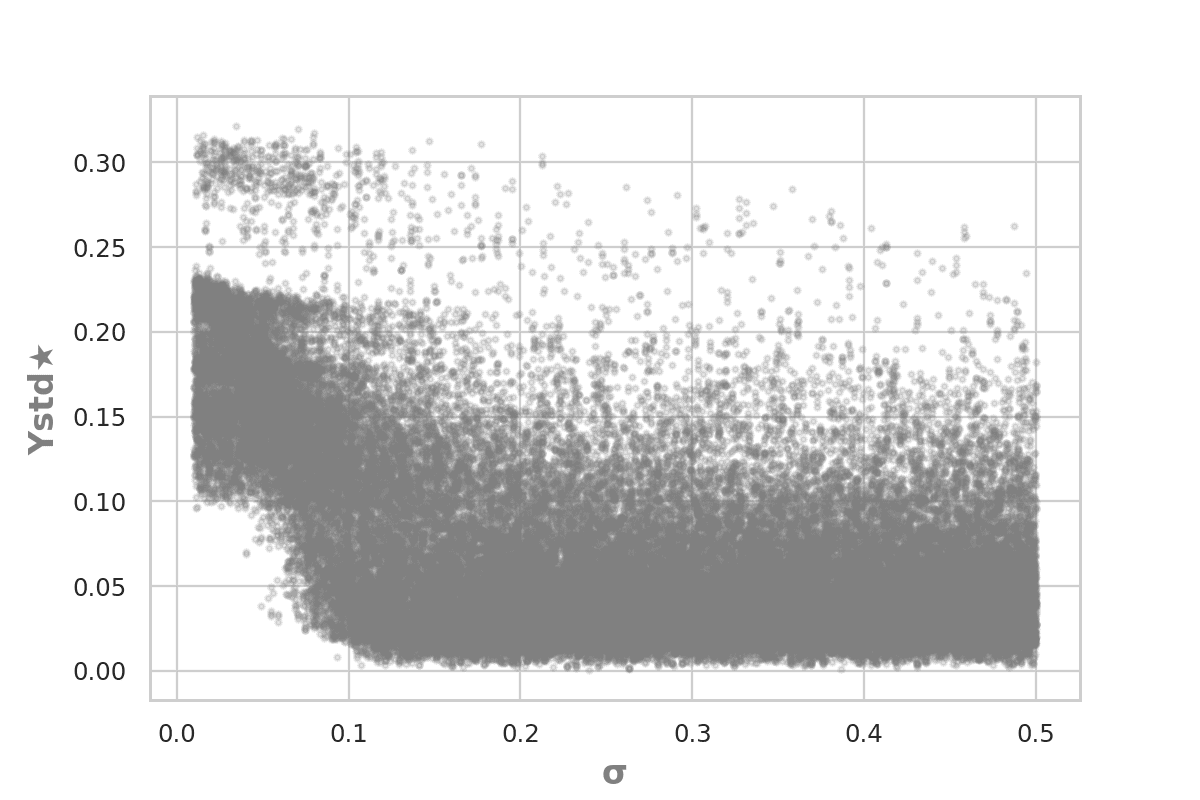}
     \end{subfigure}
     \caption{Scatterplots for the parameters with highest impact. Each dot corresponds to the result obtained in a single run.}
      \label{fig:scatters}
    \end{figure}

    A sensitivity analysis, however, does not show the direction of the impact.
    Figure~\ref{fig:scatters} shows scatter plots for $Ystd$ as a function of
    the main parameters. Each point corresponds to the outcome of one
    implementation of the model for the value of the parameter at the $x$-axis
    and the observed value of $Ystd$. Similar scatter plots for $S_s$ showed
    similar behaviors and, therefore, are not shown here.

    The negative impact of \(\sigma\) in the population' opinion dispersion is
    expected: a higher \(\sigma\) means agents are easier to influence. As they
    are connected to all the other agents, the more uncertain they are, the more
    centralized the agents' mean opinions will tend to become. The plots also
    show that the exact value of \(\sigma\) seems to matter little above
    approximately 0.1. Therefore we restrict our following analysis to \(\sigma\) in the
    \((0.0, 0.1 ] \) range. The effect of \(\rho\) is also expected: the bigger
    the noise, the more dispersed the final state of the system is. The number
    of issues $n$ seems to have little influence unless it is $n=1$. When there
    is only one issue, we see scenarios where the opinions clearly move away to
    stronger polarizations, with (\(Ystd\)) around 0.3. Those cases are no
    longer observed as we have at least two issues. That is probably an artifact of initial conditions.
    As initial conditions were randomly drawn, extreme values in all issues become more and more rare
    with increasing number of issues $n$ and, therefore, there are very few agents to push others to the more extreme regions.

\begin{figure}[H]
  \centering
       \begin{subfigure}[b]{0.48\textwidth}
      \includegraphics[width=\textwidth]{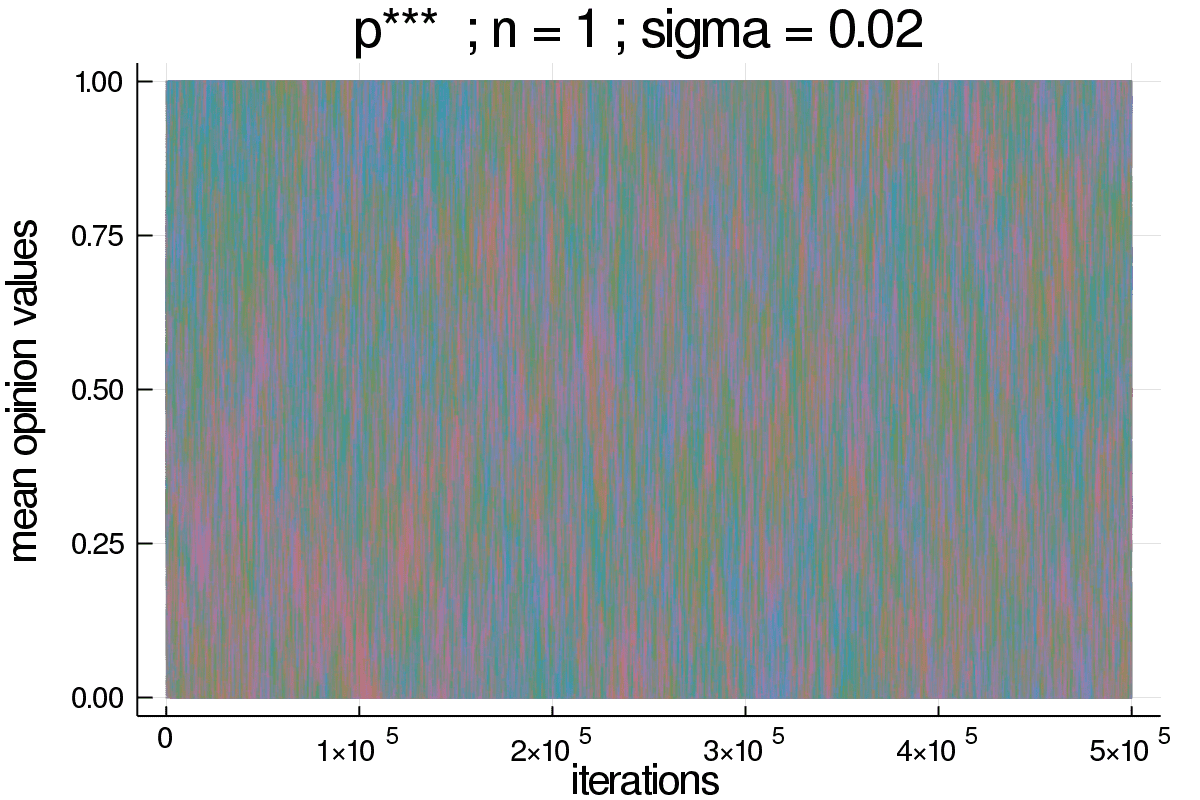}
     \end{subfigure}
    \begin{subfigure}[b]{0.48\textwidth}
      \includegraphics[width=\textwidth]{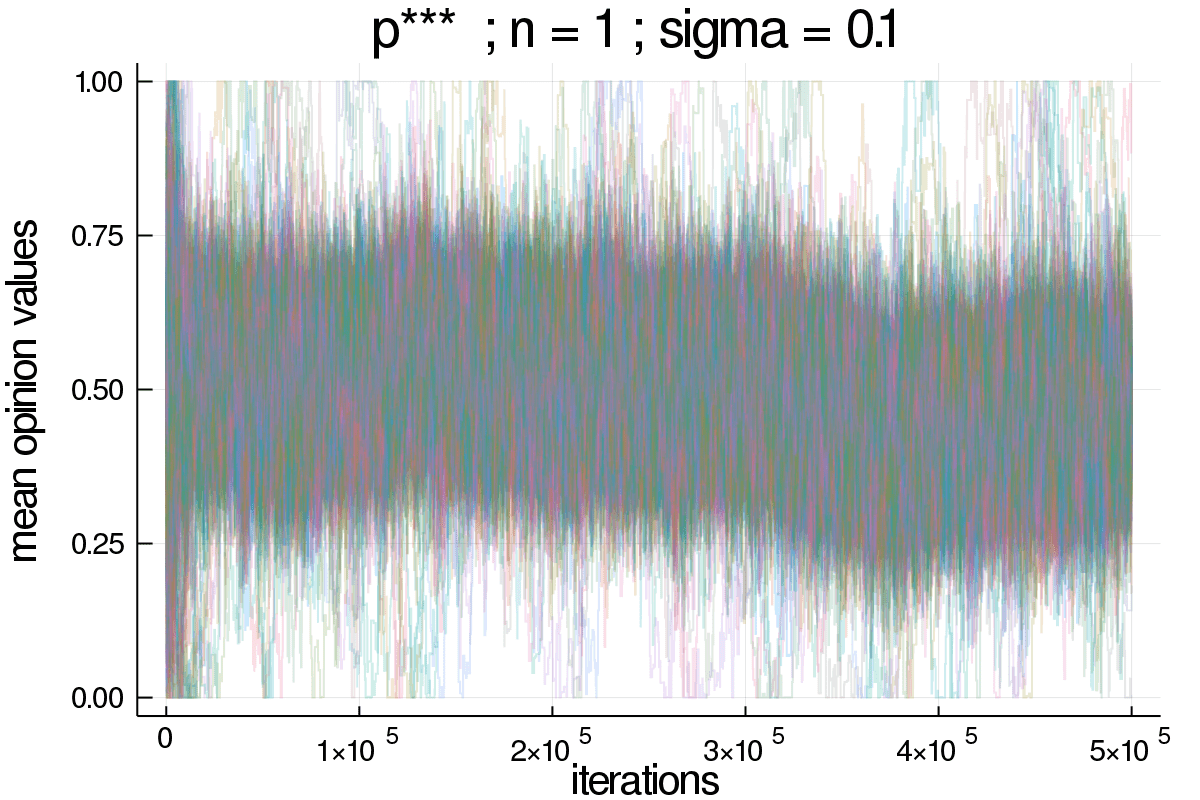}
    \end{subfigure}     
              \begin{subfigure}[b]{0.48\textwidth}
      \includegraphics[width=\textwidth]{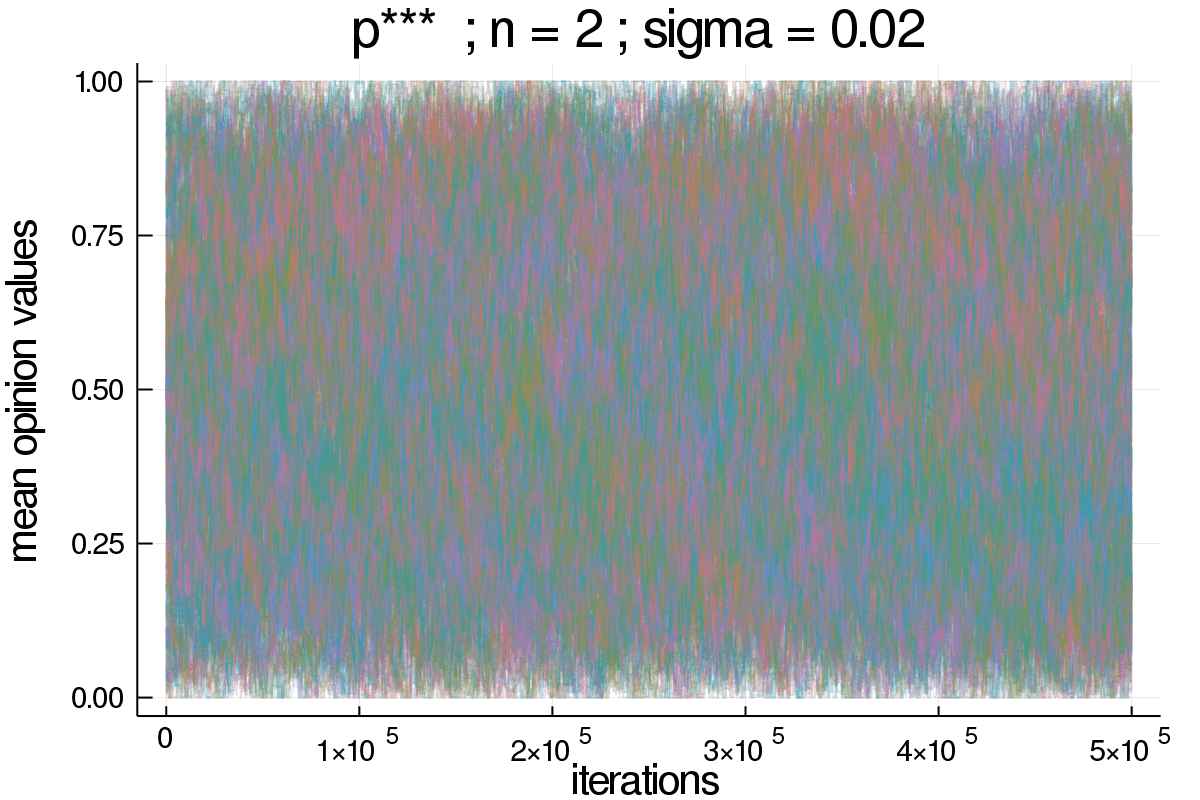}
     \end{subfigure}
    \begin{subfigure}[b]{0.48\textwidth}
      \includegraphics[width=\textwidth]{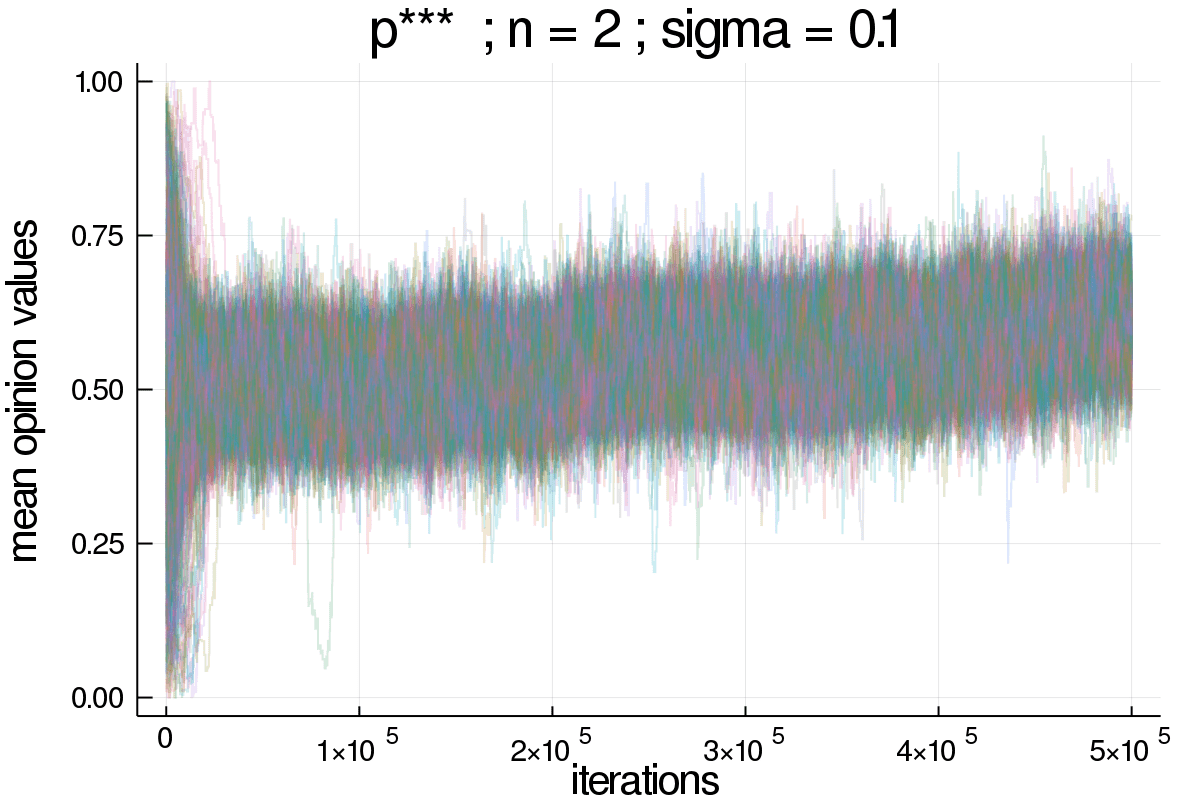}
    \end{subfigure}
     \caption{Time series for the parameterization: \(\rho = 0.05, N = 500,
       p\_intran = 0.0, n  = 1 \) or  $2$}
      \label{fig:tseries1}
    \end{figure}

    Now we have a general picture, looking at
    typical behaviors for specific parameter values helps us understand the model better. We ran cases where we kept
    \(\rho = 0.05 \) ; \(N = 500\) ; \(p\_intran = 0.0\) fixed, for 500.000
    iterations, and test combinations of $\sigma = (0.02, 0.04, 0.1)$ and $ n =
    (1,5,10)$. The time series at Figure \ref{fig:tseries1} show the time evolution
    of the opinions of all $500$ agents for a typical run of the
    the $p{***}$ case. Only results for $p{***}$ are shown here because the time
    series for $p^*$ and $p{**}$ are visually almost identical to the ones in
    the figures. We see that larger values \(\sigma\) seem to lead opinions towards
    the center: the bigger the \(\sigma\) is, the more agents will tend to move
    closer to the mean.

\begin{figure}[H]
      \centering
      \begin{subfigure}[b]{0.48\textwidth}
        \includegraphics[width=\textwidth]{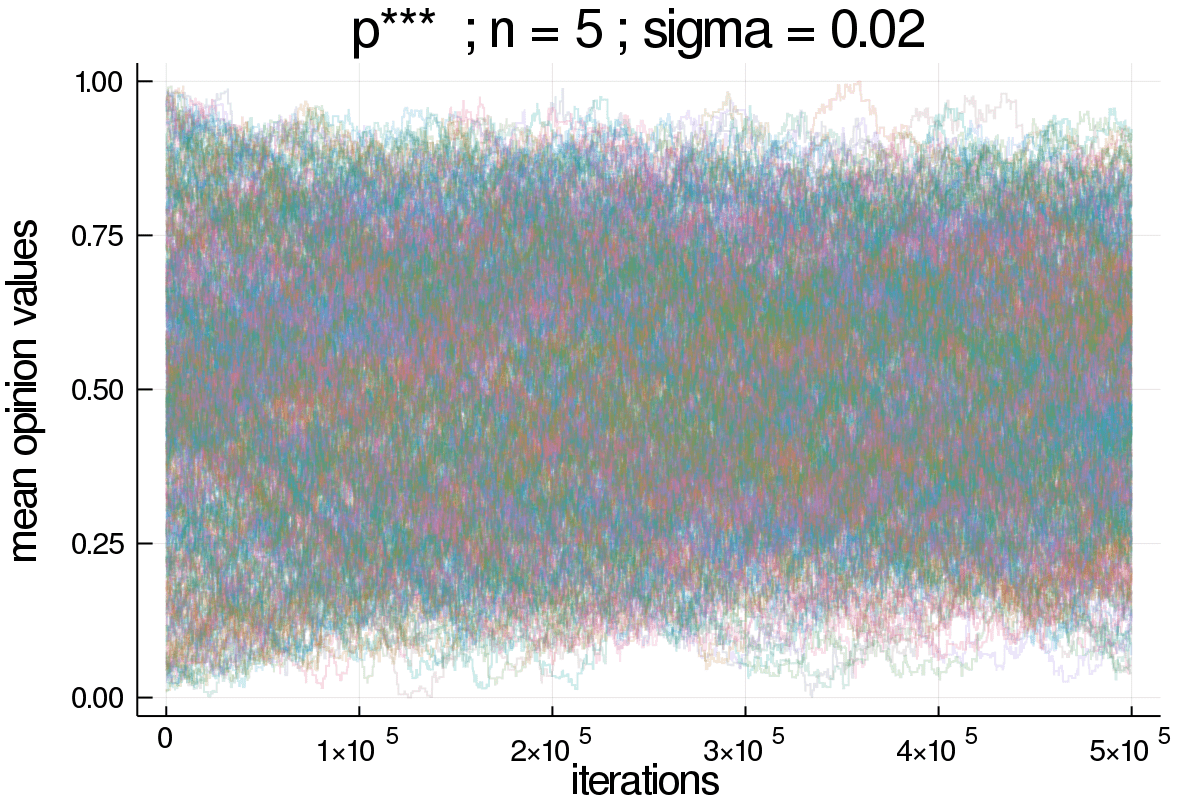}
      \end{subfigure}
      
      \begin{subfigure}[b]{0.48\textwidth}
        \includegraphics[width=\textwidth]{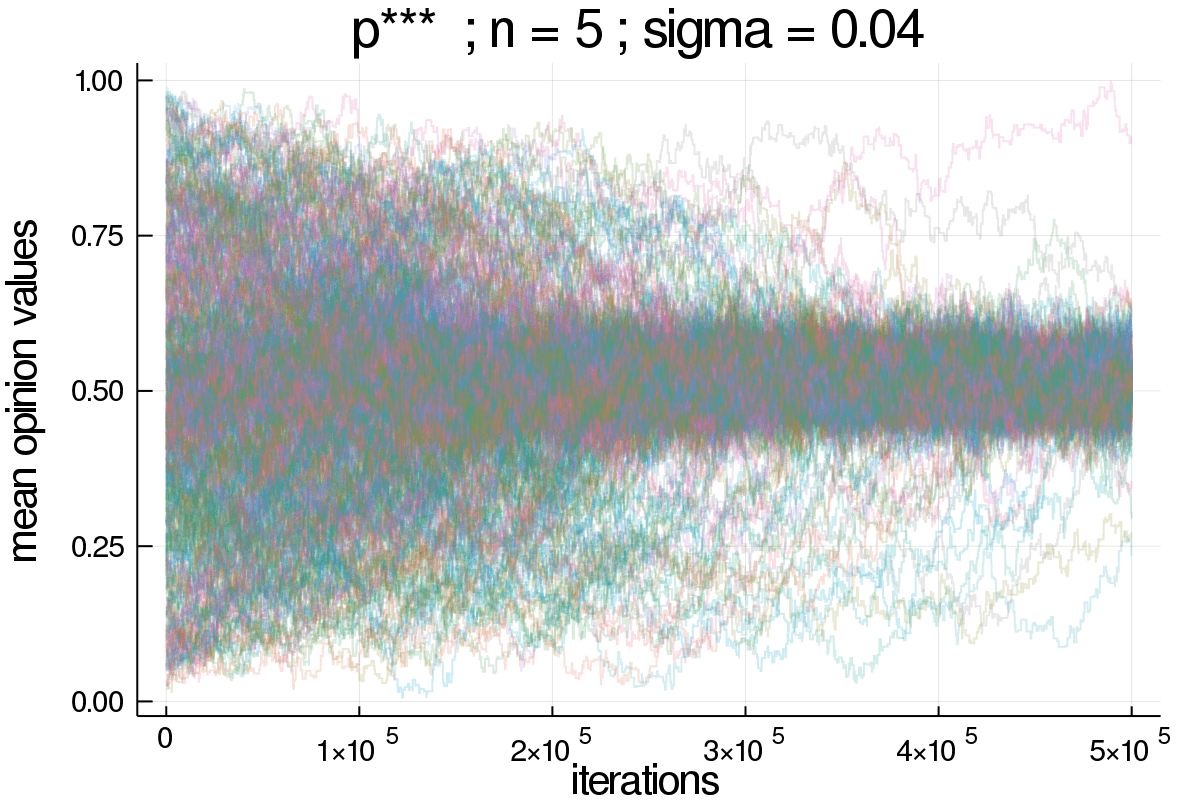}
      \end{subfigure}
      
      \begin{subfigure}[b]{0.48\textwidth}
        \includegraphics[width=\textwidth]{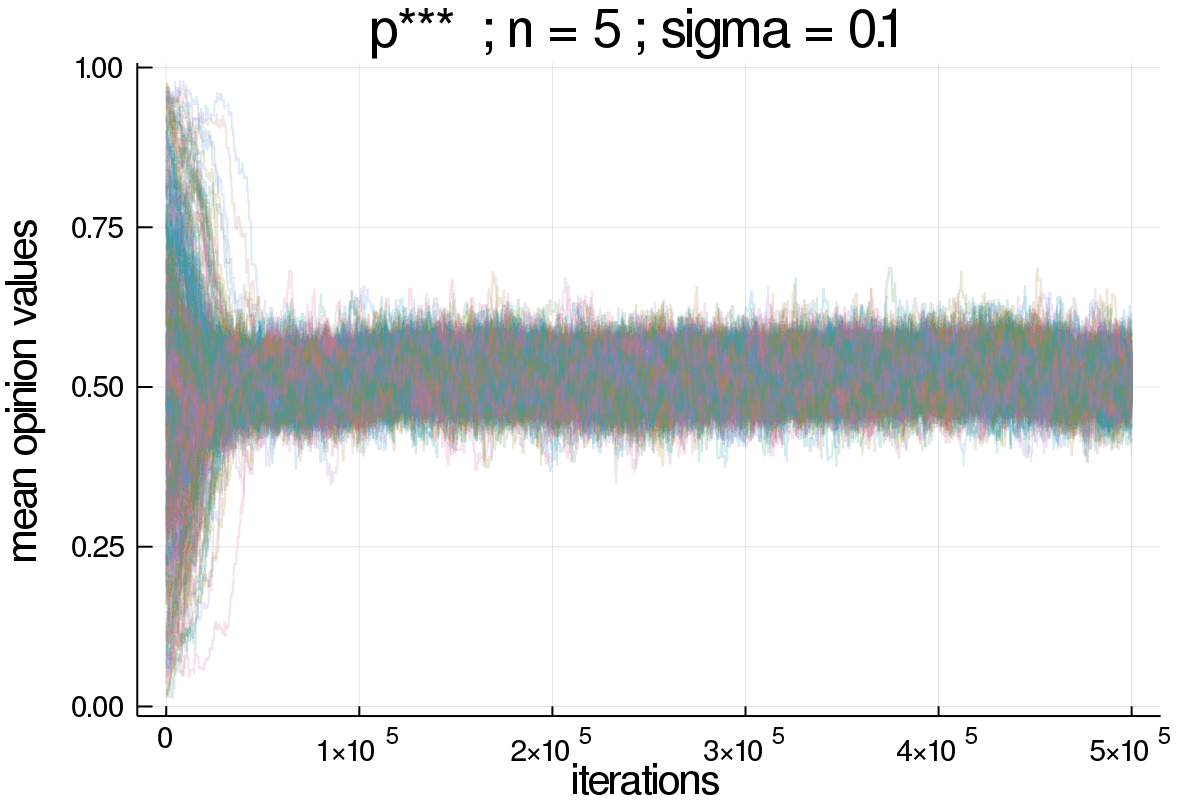}
      \end{subfigure}
      \caption{Time series for the parameterization: \(\rho = 0.05, N = 500,
        p\_intran = 0.0\)}
            \label{fig:tseries2}
          \end{figure}

    Figure \ref{fig:tseries2} shows a similar time evolution series for $n=5$
    issues. Here, even for small values of $\sigma$ we observe the opinions tend
    to avoid the more extreme values. As commented before, that seems to be an artifact of initial
    random draws. While it is easy to draw the most extreme values with only one
    draw, for $n=5$, as we are close to the end of the range, no values outside
    that range are possible. So, the most extreme values require that, at the
    start, agents should have all their five issues drawn as extreme. As that is
    rare, those few that do start there tend to be attracted to still extreme
    positions, but a little less so. However, there actually seems to be a stronger
    tendency towards more central opinions. That can be observed as the increase in $\sigma$
    leads to a position closer to consensus than we observed in Figure \ref{fig:tseries1} for the $n=1$ case.

\begin{figure}[H]
      \centering
      \begin{subfigure}[b]{0.48\textwidth}
      \includegraphics[width=0.9\textwidth]{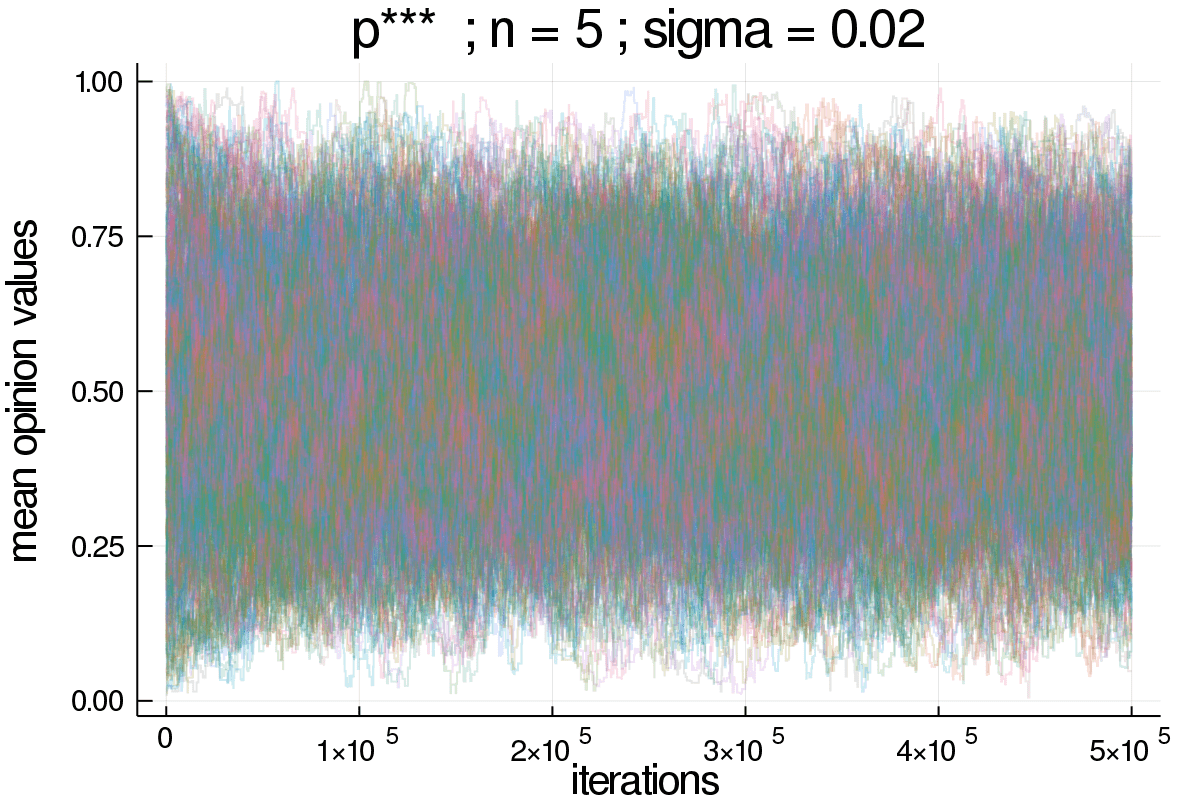}
    \end{subfigure}
    \begin{subfigure}[b]{0.48\textwidth}
      \includegraphics[width=0.9\textwidth]{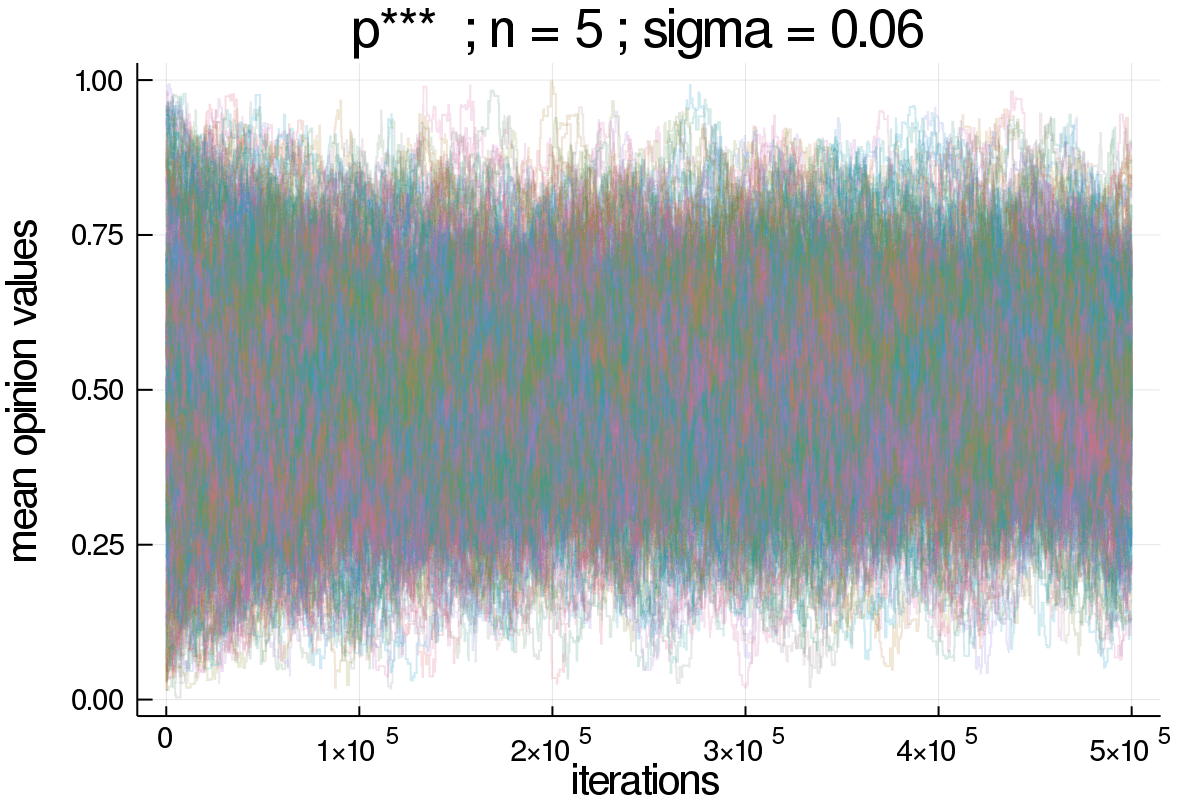}
    \end{subfigure}
      \begin{subfigure}[b]{0.48\textwidth}
      \includegraphics[width=0.9\textwidth]{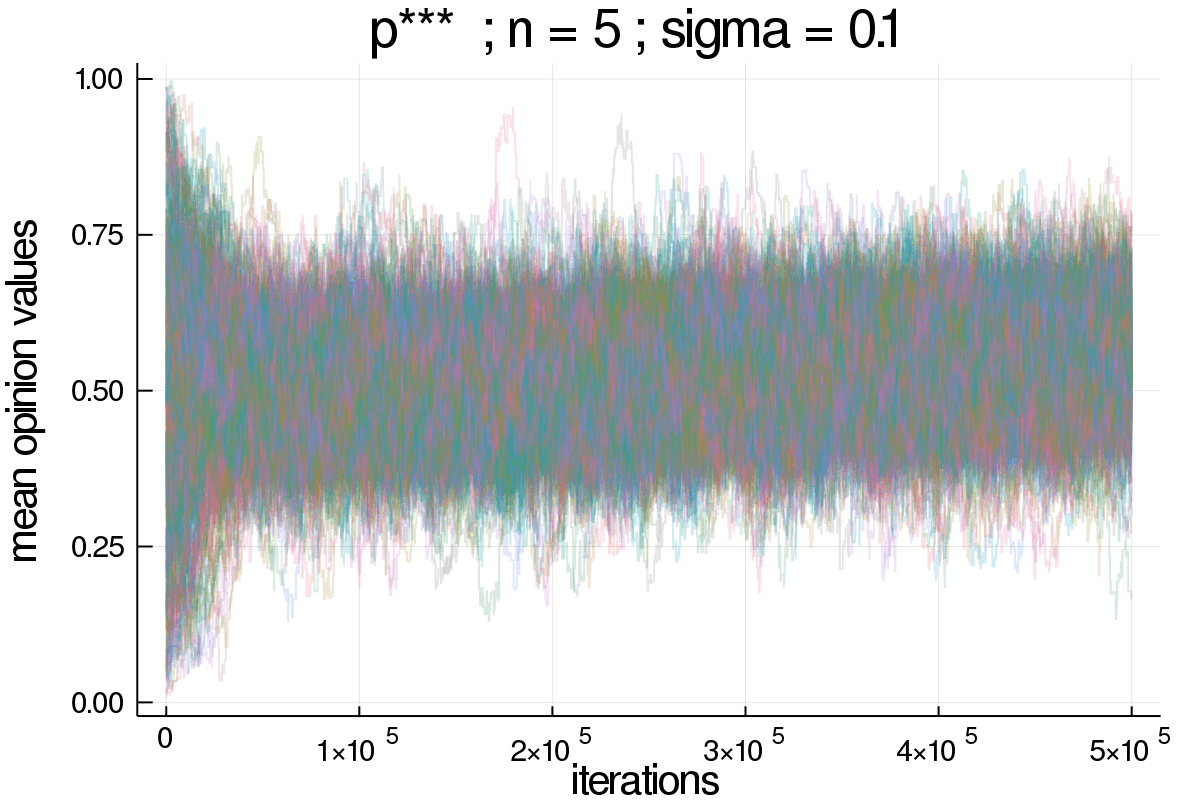}
    \end{subfigure}
    
      \caption{Time series for the parameterization: \(\rho \approx 0.12, N =
        500, p\_intran = 0.0 \)}
      \label{fig:tseries3}
    \end{figure}

    A possible reason we need to check is the fact we are  measuring the mean opinion values \( x_i \).
    As \(\rho\) changes a single \(o_i\) at each iteration, that means a
    higher \(n\) should imply a lesser impact of \(\rho\) on the mean opinion of the
    agent. That could explain why, for a larger $n$, consensus might come easier. To test if that was what was actually happening, we ran the same scenarios but increasing \(\rho\) with \(n\), such that  \(\rho_2
    = \sqrt{n} * \rho_1 \). The effects of increasing \(\rho\) that way in the case of Figure \ref{fig:tseries2} can be seen at Figure \ref{fig:tseries3}. A comparison with Figure \ref{fig:tseries1} shows a larger spread around the consensus, suggesting there might be other effects here other than random fluctuations.

    Heretofore we've tested parameterizations with noise. To verify the
    influence of the existence of noise, we also ran cases with \(\rho\) close to
    zero,that is \(\rho = 10^{-5} \). The most obvious difference we can see is
    that the population mean opinion values converge to well defined values. In
    parameter combinations in which \(\sigma = 0.1\) the tendency is convergence
    to values close to 0.5. An interesting distinction between the cases in this
    parameterization is that \(p^{**}\) and \(p^{***}\) always converge to 0.5,
    independently of the number of issues. Alternatively, in the \(p^{*}\) case
    this happens when \(n=1\), but when we have \(n=5\) or \(10\) there are
    other values of convergence, more as we increase \(n\), even though the
    centralizing tendency remains.
    \begin{figure}[H]
      \centering
      \begin{subfigure}[b]{0.48\textwidth}
        \includegraphics[width=\textwidth]{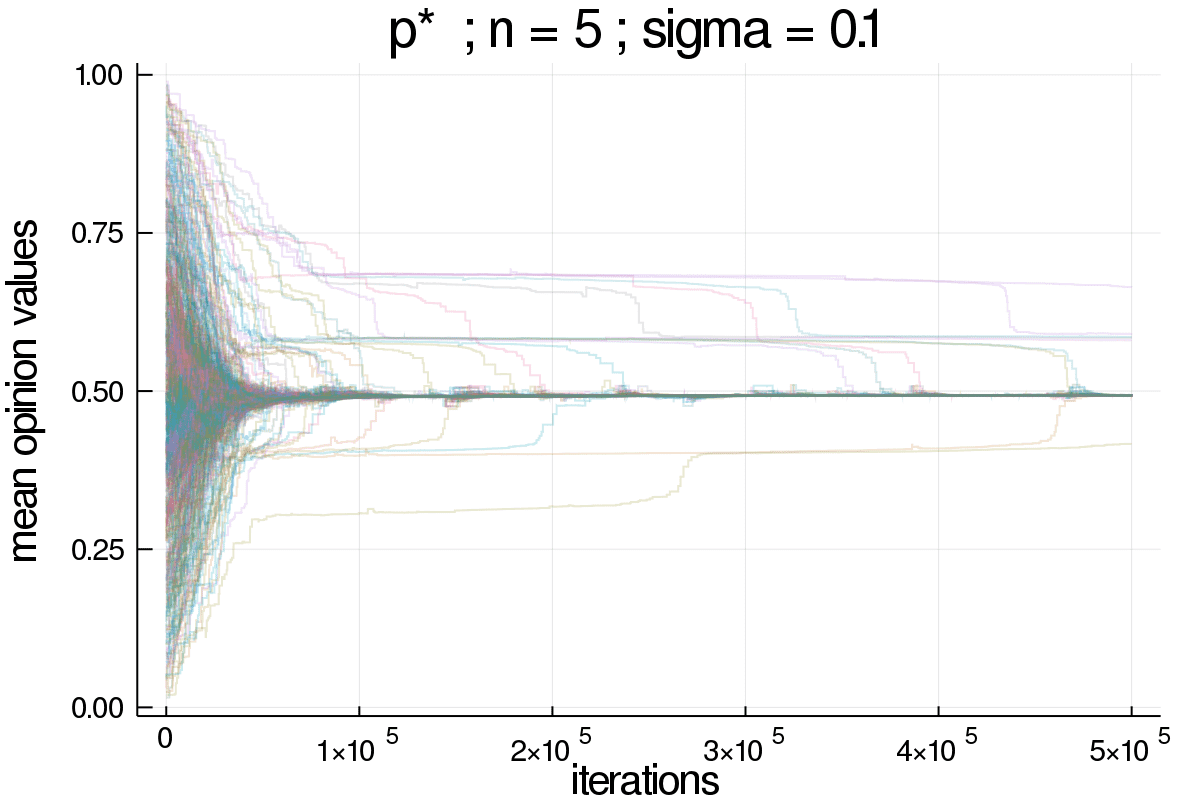}
      \end{subfigure}
      \begin{subfigure}[b]{0.48\textwidth}
        \includegraphics[width=\textwidth]{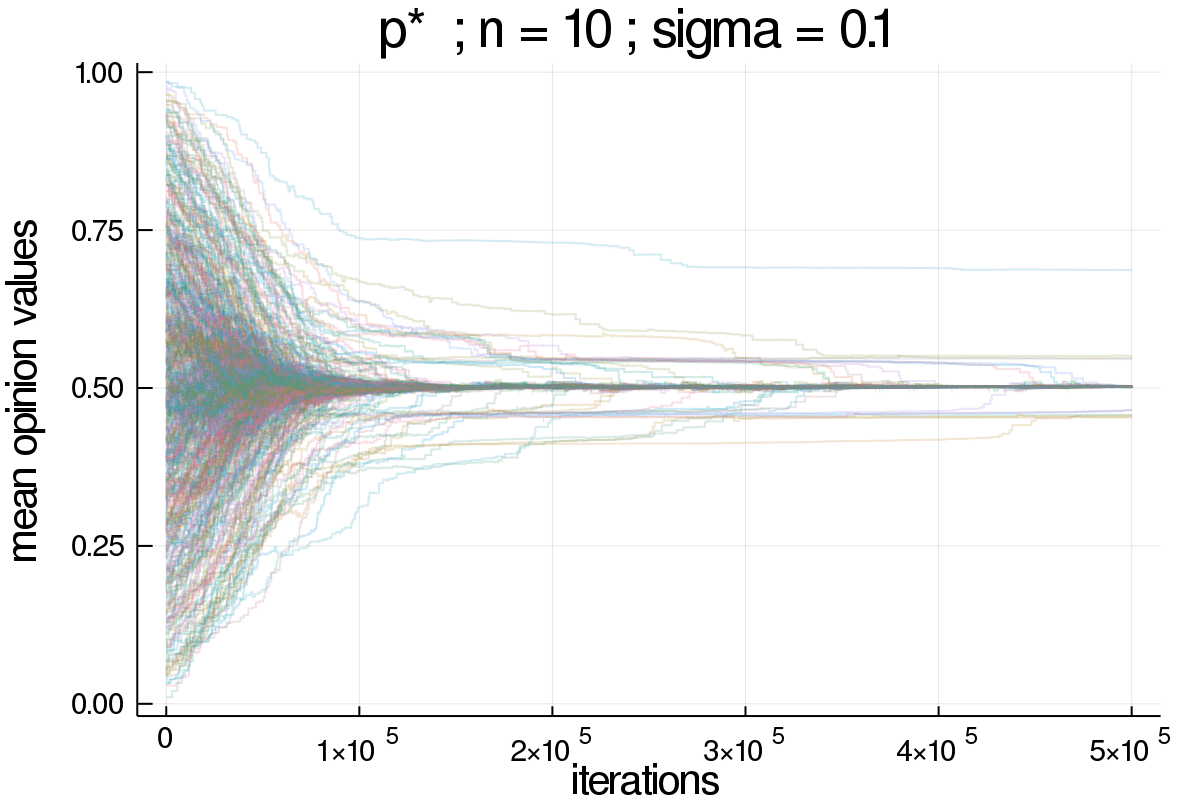}
      \end{subfigure}
      \begin{subfigure}[b]{0.48\textwidth}
        \includegraphics[width=\textwidth]{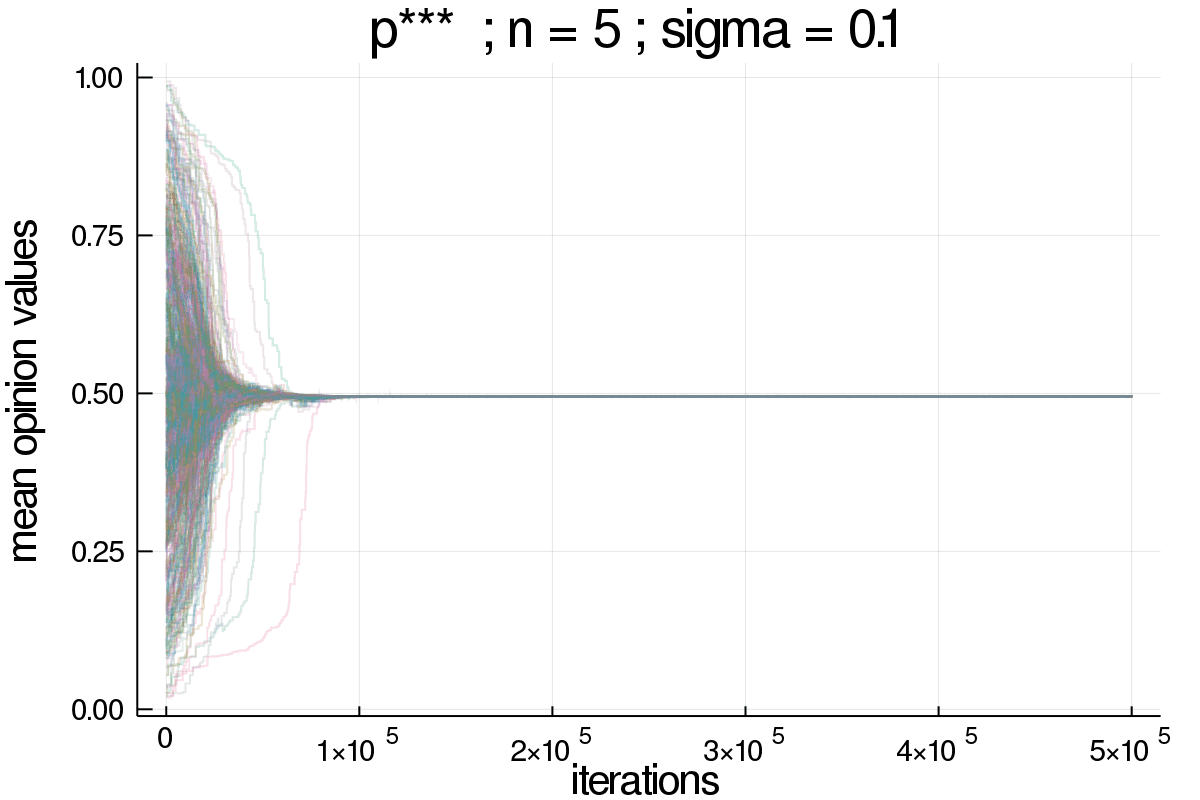}
      \end{subfigure}
            \begin{subfigure}[b]{0.48\textwidth}
        \includegraphics[width=\textwidth]{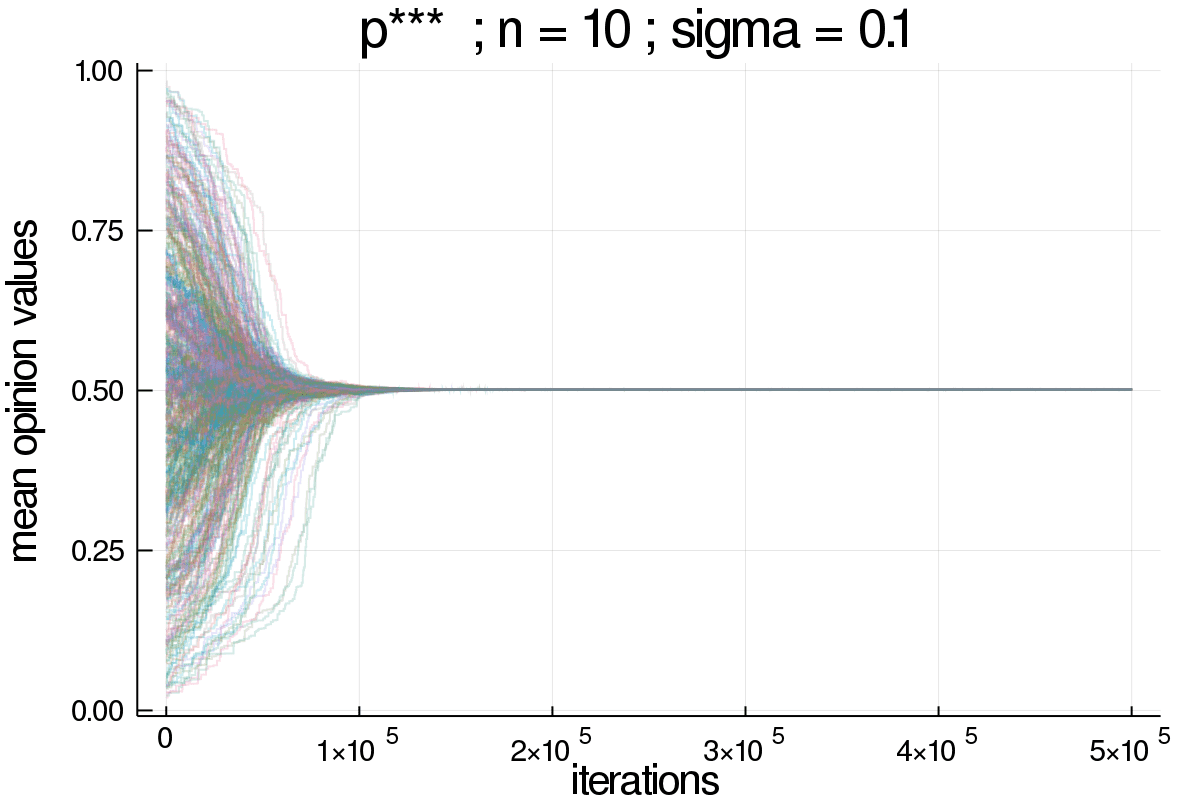}
      \end{subfigure}
      \caption{Time series for the parameterization: \(\rho = 1e-5, N =
        500, p\_intran = 0.0 \)}
      \label{fig:tseries4}
    \end{figure}
    
\begin{figure}[H]
      \centering
      \begin{subfigure}[b]{0.48\textwidth}
        \includegraphics[width=\textwidth]{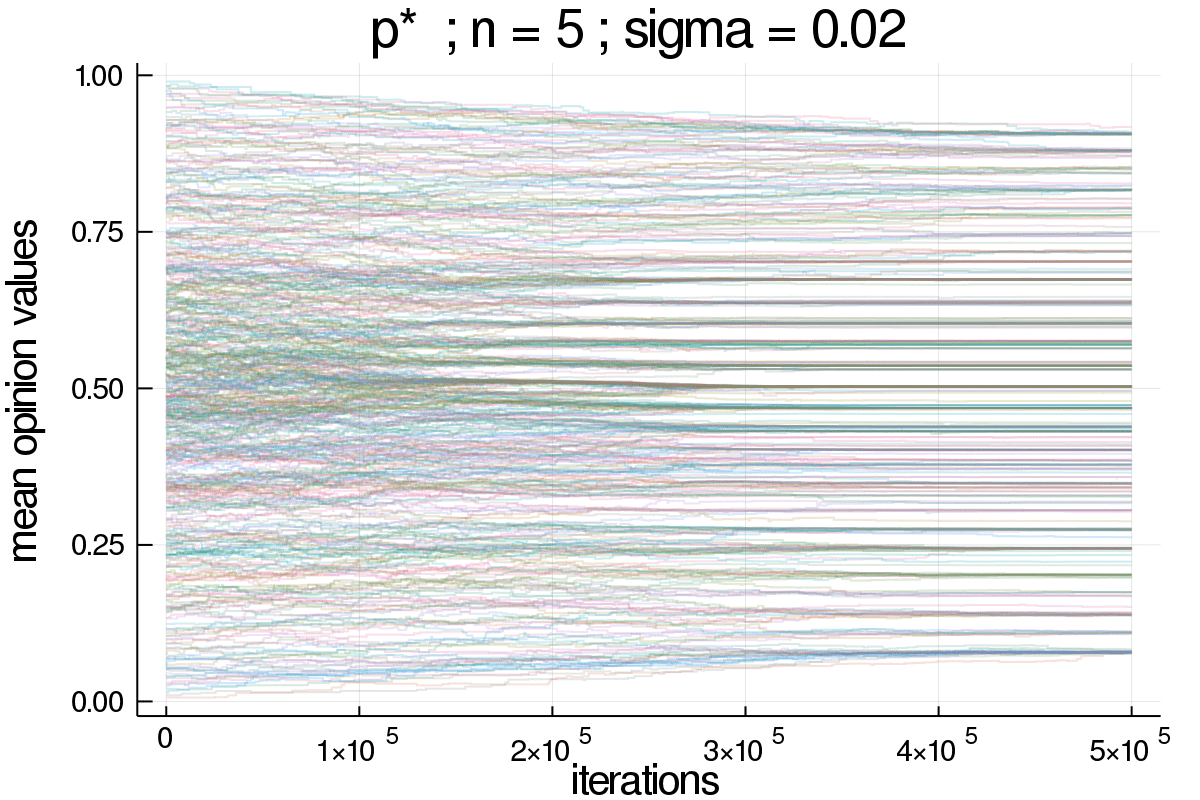}
      \end{subfigure}
      \begin{subfigure}[b]{0.48\textwidth}
        \includegraphics[width=\textwidth]{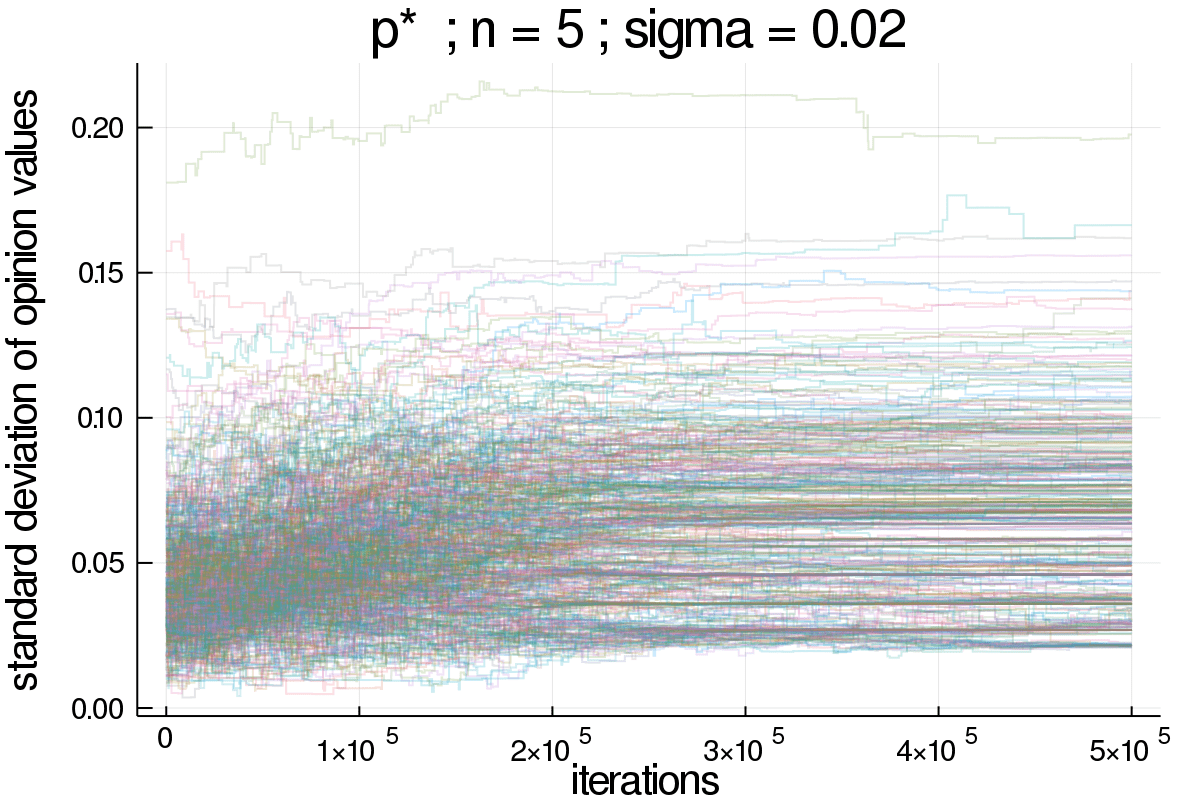}
        \end{subfigure}
      \begin{subfigure}[b]{0.48\textwidth}
        \includegraphics[width=\textwidth]{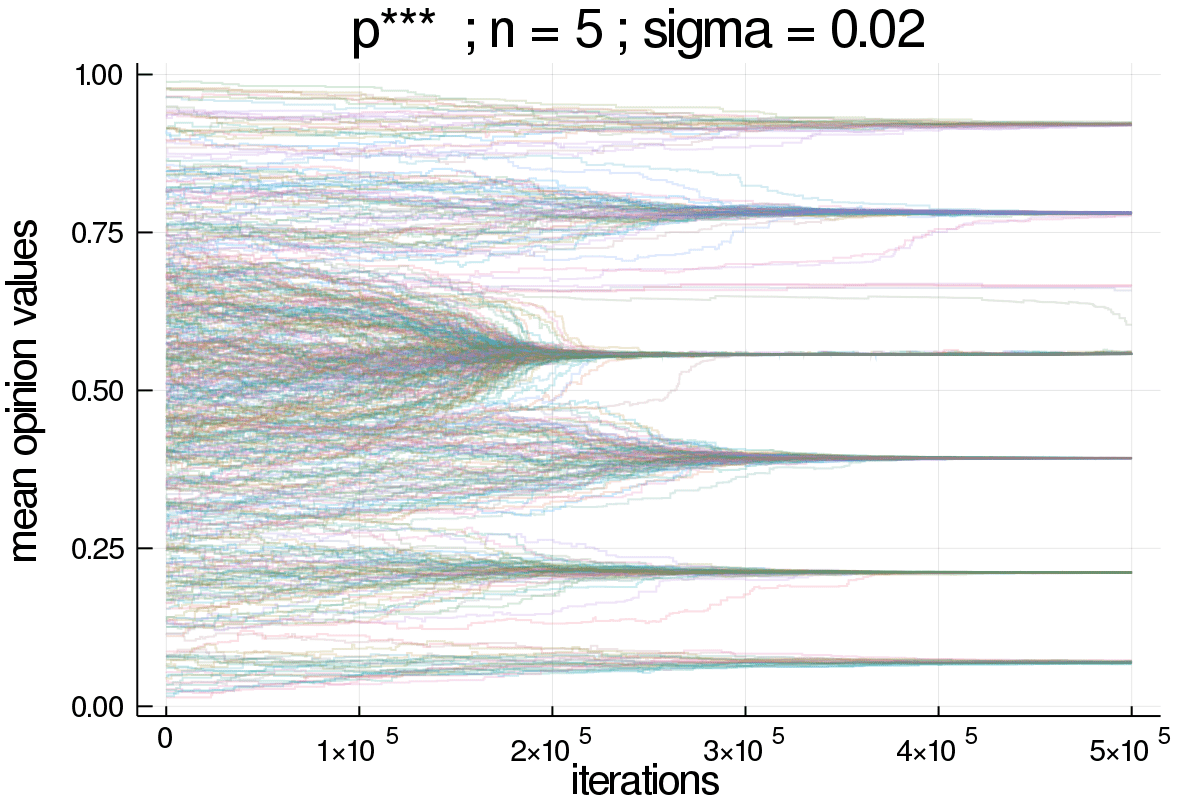}
      \end{subfigure}
      \begin{subfigure}[b]{0.48\textwidth}
        \includegraphics[width=\textwidth]{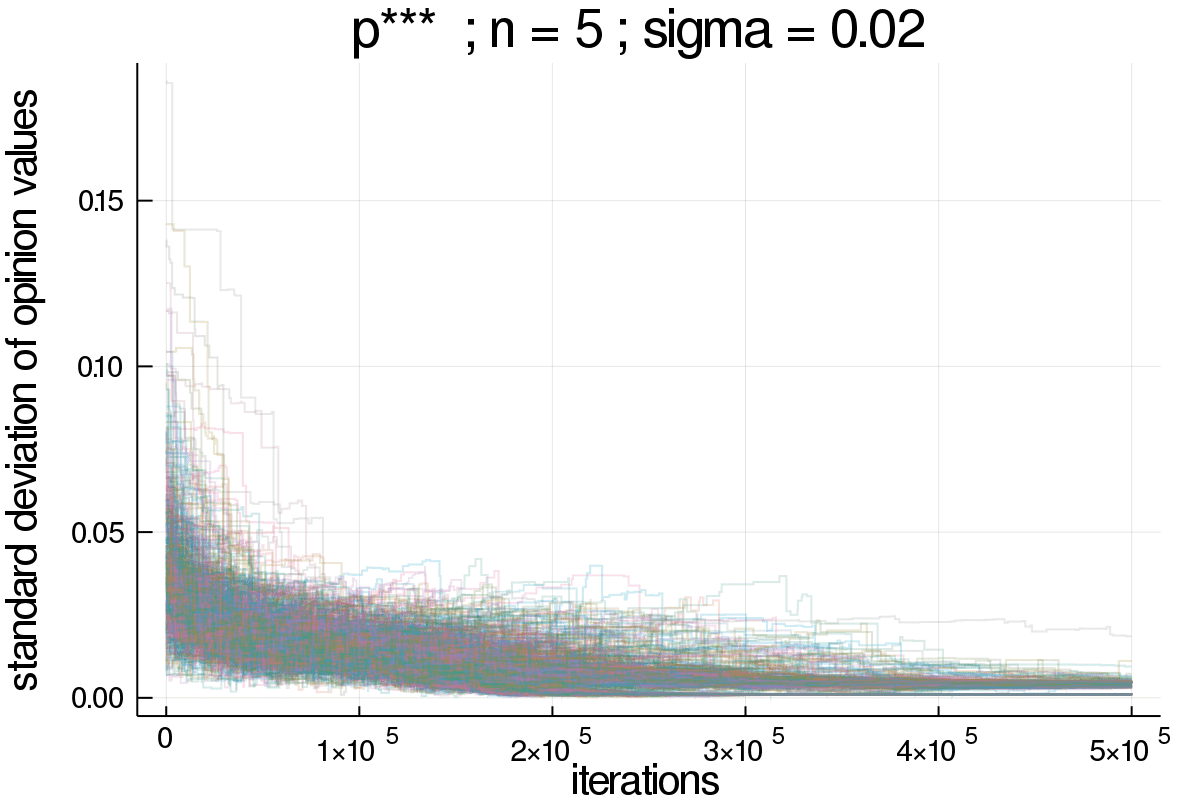}
      \end{subfigure}
      \caption{Time series for the parameterization: \(\rho = 1e-5, N =
        500, p\_intran = 0.0 \)}      
  \label{fig:tseries5}
    \end{figure}

   When \(\sigma\) is around \(0.02\) or \(0.04\), we observe another difference
    between cases: \(p^{*}\) has more convergence values than \(p^{**}\) and
    \(p^{***}\). Figure \ref{fig:tseries5} illustrates this system behavior for $\sigma=0.02$. We can also see that the standard deviations $s_i$ are very different for the \(p^{*}\) and \(p^{***}\) cases. While for \(p^{*}\), the values of $s_i$ range from small values around 0.02 to much larger values, up tp 0.2, in the \(p^{***}\) scenario we see that $s_i$ tends to diminish with time, the majority of the observed values getting close to zero.

    The reason for that lies in how \(\Delta\) (that controls how much an agent will trust others) is calculated in each case: in the \(p^{**}\)
    and \(p^{***}\) cases the update rules make use of mean opinion values. That introduces a tendency for distinct issues opinions to move towards each other. The \(p^{*}\) update rule, however, works with
    single issue opinions. As the issues are uncoupled, that allows the agents to keep an ideologically more diverse profile.
    
\begin{figure}[H]
      \centering
      
      \begin{subfigure}[b]{0.49\textwidth}
        \includegraphics[width=\textwidth]{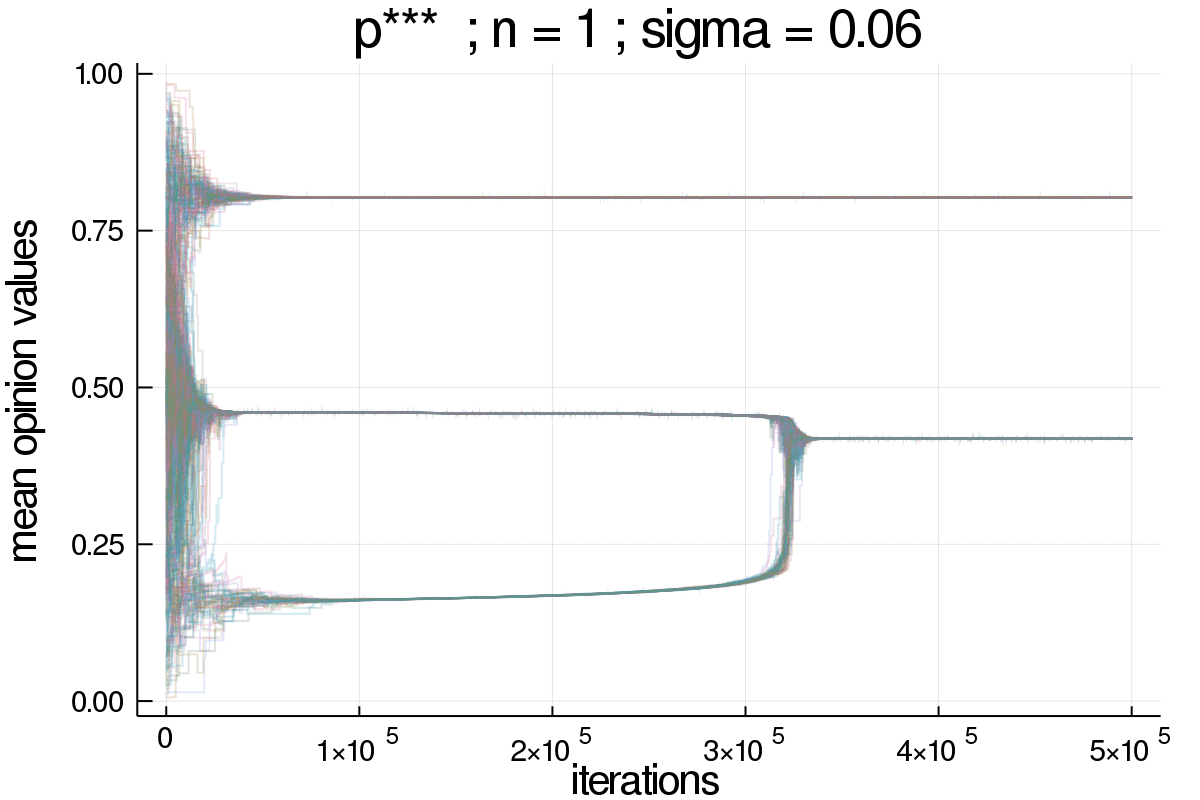}
      \end{subfigure}
      
      \begin{subfigure}[b]{0.49\textwidth}
        \includegraphics[width=\textwidth]{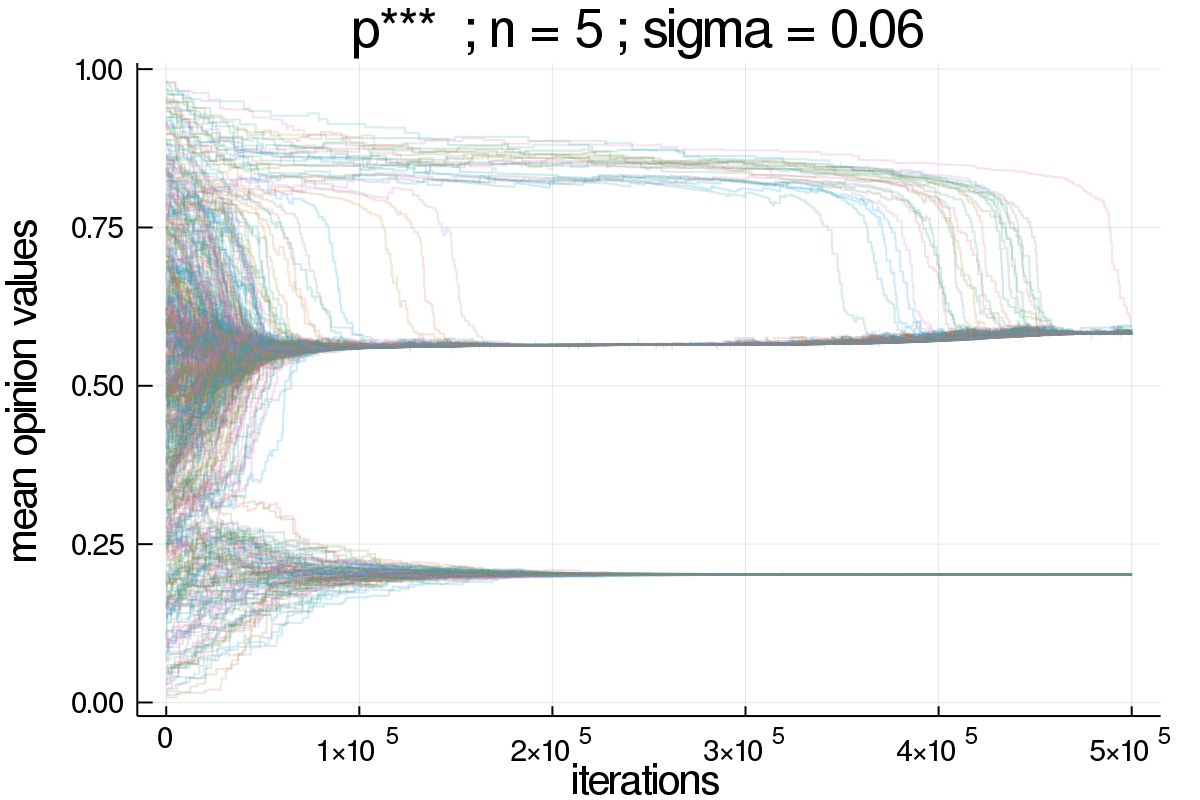}
      \end{subfigure}
            \begin{subfigure}[b]{0.49\textwidth}
        \includegraphics[width=\textwidth]{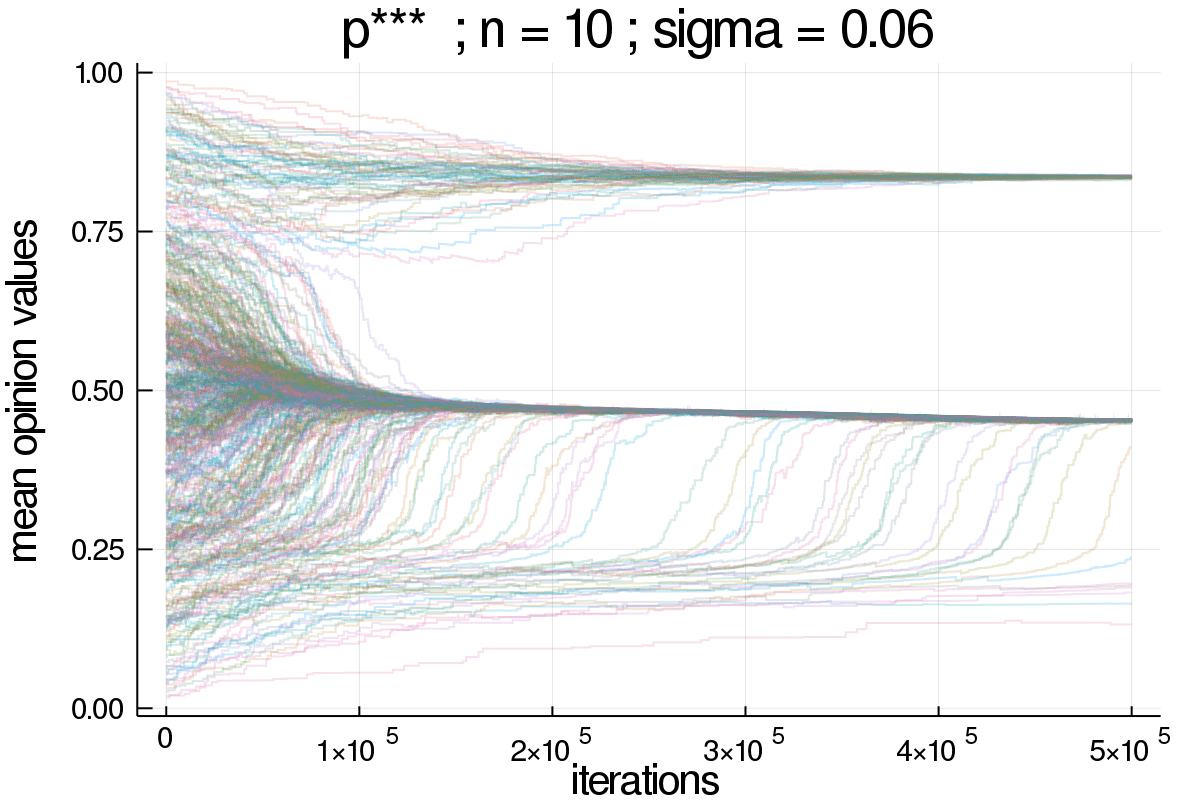}
      \end{subfigure}
      \caption{Time series for the parameterization: \(\rho = 1e-5, N = 500,
        p\_intran = 0.0 \).}
  \label{fig:tseries6}
    \end{figure}

    Another impact of the number of issues, as shown in Figure
    \ref{fig:tseries6}, is that a higher \(n\) leads to a longer time for the
    convergence. The reason is that we're only changing one
    opinion by iteration, so naturally a higher \(n\) means the agents will take
    longer to be influenced. The relationship here is roughly linear such that
    the plot the region at \(5 \times 10^5\) iterations when \(n = 10\) is very
    similar to the corresponding region at \(0.5 \times 10^5\) when \(n = 1\).
    
    \section{Conclusions}

    Here, we have extended a previous model for continuous opinions
    \cite{martins08c} to include the case where agents have opinions on $n$ issues. We explored
    the impact of the main parameters on the final results. The \(\sigma\) parameter, that defines how much
    each agent trusts other with distant opinions and plays the same role threshold values play in Bounded Confidence models, was observed to be a major influence in the outcome of the simulations. The number of issues $n$ as well as the amount of noise $\rho$ were also shown by our sensitivity analysis to have important effects on the outcomes, while the initial general amount of trust $p$ as well as the number of agents $N$ seemed to matter little.
    
    As the model is based on an update rule where opinions always move towards agreement (even if by a negligible amount), the tendency to central opinions is expected. That is especially true for the average values $x_i$. Since they are an average of several opinions, a strong tendency to central values is expected from simple statistical considerations. In order to see evidence of cases where the final opinions show a splitting between several final values, we do have to look at the individual opinions on each issue, $o_{ik}$. It is interesting to see that, at least for the regular trust function $p^*$, the final distributions for  $o_{ik}$ show that specific opinions might become quite spread over the range of possible values. 
    
    Testing different forms of trust functions also provided interesting results. A Bayesian analysis of the problem suggests the rational choice would be just comparing the opinions of the agents on the specific issue. That is, how much each agent $i$ trusts an agent $j$ should be a function of the distance between their opinions on the subject $k$ they are debating, that is, $o_{ik}-o_{jk}$. However, as humans do show a lot of ideologically motivated reasoning  \cite{mercier11a,merciersperber11a,kahanetal11}, it makes sense to change how trust is calculated to a situation that is more compatible with experiments. Two possibilities were considered here.  In the first one,   (\(  p^{**}\), the  trust was dependent on the distance between the agent average estimates over all issues, that is, the trust function was a function of $x_{i}-x_{j}$. The second possibility, \( p^{***}\)), considered that agent $i$ was observing $j$ opinion only on issue $k$ and, therefore, could only compare that value to its own average, that is the trust function was a function of $o_{ik}-x_{j}$.
   
	Compared to the non-ideological case, both ideological trust functions showed a tendency that each agent would have it own opinions show less diversity over the set of issues. This could be easily observed by the larger proportion of smaller values of $s_i$. That effect was particularly stronger for  \( p^{***}\)). As the issues were always treated as independent, that tendency does correspond to observing some irrational consistency~\cite{jervis76a}. The tendency to consistency was not perfect, though, as large values of $s_i$ were also observed as much more common than those in the initial distributions. That suggests that, while the mechanism of ideological trust might play an important role in the existence of the irrational consistency effect, it probably can not account for the whole of it. Other effects, such as confirmation biases, are also probably important to describe the whole effect.

\section{Acknowledgement}
The author would like to thank Funda\c{c}\~ao de Amparo \`a Pesquisa do Estado de S\~ao Paulo (FAPESP), for the support to this work, under grant 2014/00551-0.

\bibliography{poodl-paper}

\end{document}